\PassOptionsToPackage{table}{xcolor} %
\documentclass[sigconf]{acmart}

\usepackage{makecell}
\usepackage{balance}

\usepackage[utf8]{inputenc}
\usepackage{enumitem}
\usepackage{arydshln} %
\usepackage{pbox}
\newlist{inlinelist}{enumerate*}{1}
\setlist*[inlinelist,1]{%
  label=(\roman*),
}

\theoremstyle{definition}
\newtheorem{definition}{Definition}[section]

\title[On Natural Language User Profiles for Transparent and Scrutable Recommendation]{On Natural Language User Profiles for\\Transparent and Scrutable Recommendation}

\author{Filip Radlinski}
  \affiliation{%
    \institution{Google}
    \city{London}
    \country{UK}}
  \email{filiprad@google.com}

\author{Krisztian Balog}
  \affiliation{%
    \institution{Google}
    \city{Stavanger}
    \country{Norway}}
  \email{krisztianb@google.com}

\author{Fernando Diaz}
  \affiliation{%
    \institution{Google}
    \city{Montreal}
    \country{Canada}}
  \email{diazfernando@google.com}

\author{Lucas Dixon}
  \affiliation{%
    \institution{Google}
    \city{Paris}
    \country{France}}
  \email{ldixon@google.com}

\author{Ben Wedin}
  \affiliation{%
    \institution{Google}
    \city{Cambridge}
    \country{USA}}
  \email{wedin@google.com}

\copyrightyear{2022}
\acmYear{2022}
\setcopyright{rightsretained}

\acmConference[SIGIR '22]{Proceedings of the 45th International ACM SIGIR Conference on Research and Development in Information Retrieval}{July 11--15, 2022}{Madrid, Spain.}
\acmBooktitle{Proceedings of the 45th Int'l ACM SIGIR Conference on Research and Development in Information Retrieval (SIGIR '22), July 11--15, 2022, Madrid, Spain}
\acmDOI{10.1145/3477495.3531873}
\acmISBN{978-1-4503-8732-3/22/07}

\usepackage{etoolbox}
\makeatletter
\patchcmd{\maketitle}{\@copyrightpermission}{
   \begin{minipage}{0.3\columnwidth}
     \href{http://creativecommons.org/licenses/by/4.0/}{\includegraphics[width=0.90\textwidth]{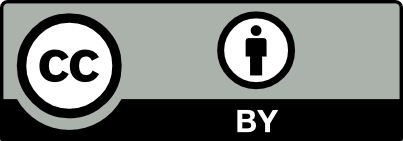}}
   \end{minipage}\hfill
   \begin{minipage}{0.7\columnwidth}
     \href{http://creativecommons.org/licenses/by/4.0/}{This work is licensed under a Creative Commons Attribution International 4.0 License.}
   \end{minipage}

   \vspace{5pt}
}{}{}

\makeatother

\begin{document}
\fancyhead{}

\begin{abstract}
Natural interaction with recommendation and personalized search systems has received tremendous attention in recent years.
We focus on the challenge of supporting people's understanding and control of these systems and explore a fundamentally new way of thinking about representation of knowledge in recommendation and personalization systems.
Specifically, we argue that it may be both desirable and possible for algorithms that use natural language representations of users' preferences to be developed. 
We make the case that this could provide significantly greater transparency, as well as affordances for practical actionable interrogation of, and control over, recommendations. Moreover, we argue that such an approach, if successfully applied, may enable a major step towards systems that rely less on noisy implicit observations while increasing portability of knowledge of one's interests.

\end{abstract}



\maketitle

\section{Introduction}
\label{sec:intro}

Personalized search and recommendation systems rely on a user representation to evaluate the match between a person and candidate results. These representations are often difficult to interpret, modify, and explain~\cite{zhang:explainable-recsys}.
Whether the information need is a standing interest (such as with movie recommendation) or acute (such as when someone has a current information need and issues a search query), personalized systems rely on some representation to identify which results have maximal value. 
For example, a common representation of a user is an embedding in a continuous vector 
space~\cite{recsys-handbook}.  As such representations can be complex, systems rarely show them to the person receiving recommendations. 
Moreover, when representations are difficult for people to interpret, even when a user can see them they are unlikely to help the user understand or control the recommendations they receive.

We contrast this common approach with a proposed future where systems represent users explicitly with \emph{scrutable} natural language (NL); where scrutable language is defined as being both short and clear enough for someone to review and edit directly. 
Ideally, these natural language statements of preferences correspond to how users would either \emph{choose to} describe their own interests
in a given context, or \emph{agree with} were this description presented to them.
We envision such representations providing transparency, and allowing control of a system's personalization.
For example, such NL user models could support people updating their representation when their preferences change, distinguishing between what they would like in the future from how they behaved in the past \cite{ekstrand:2016:better-recommendations}.  
Traditional systems can be slow to recognize and adapt to such changes.
Moreover, should a person choose to maintain a summary of their interests elsewhere, they could choose when and how to share it with other services they interact with. 
In terms of recommender quality, recent research suggests that even a simplistic natural language summary of a user's interests can provide recommendations not far from those generated by collaborative filtering methods~\citep{Balog:2019:SIGIR}.

We present a decomposition of the difficult challenge posed by natural language user representations into two steps: (1) obtaining a reliable, complete, yet succinct natural language description of a user's interests, and (2) recommending items based (exclusively) on this description.
The first task, {\bf \emph{scrutable user model generation}}, involves taking input similar to that used by systems today (be these explicit user preferences, ratings over sets of items, or past interactions) to generate a length-bounded natural language summary that may be scrutinized by the user. 
Most directly related to explainable recommendation~\cite{zhang:explainable-recsys}, the natural language user model \emph{describes what constitutes relevant results} and can also be utilized to generate \emph{genuine explanations for specific item recommendations}. 
The second task, {\bf \emph{scrutable NL-based recommendation}}, then takes a natural language user model as input to assess the match between items and the user model. While most related to long-form text retrieval~\cite{Gupta:2015:FnTIR}, this task is significantly more challenging in that the long form must represent the person's interests holistically, rather than being focused on one acute need. In other words, long-form retrieval most often presents a set of conditions that must all be satisfied, while in the recommendation setting the goal is to capture the breadth of the recipient's interests (often along with context).
We also note that traditionally recommendation and personalization place no constraints on the intermediate representation, and optimize user model generation and recommendation jointly. We argue that there are transparency and scrutability benefits with a NL constraint. Given the current state of the art, it may also be valuable to consider these tasks independently, drawing on advances in large scale models for natural language understanding to accelerate progress.

There are, of course, many challenges that would need to be addressed to reach state of the art performance in personalization and recommendation while obtaining the benefits we present. Our primary goal is to expand upon these challenges, collating evidence from 
recent work suggesting the feasibility of these being solved, and ultimately motivate progress in addressing them.
While we do not expect this to yield immediate solutions, we argue they are worth investigating due to the potential larger value to society.

In summary, our main contributions are (1) a formal proposal of a fundamentally new approach to recommendation, where systems solely rely on a natural language user model that is fully transparent and controllable by the user; (2) a tractable decomposition of the approach, collating recent advances that may enable significant progress by the community; (3) a detailed presentation of specific open research challenges; (4) a qualitative feasibility study showing how recent advances in language models provide evidence towards the tractability of the problem.

\section{Overview}
\label{sec:overview}

We argue for a new philosophy for recommendation and personalization, one based on succinct natural language summaries of users' preferences rather than mathematically convenient numerical representations. 
These summaries may be inferred from implicit behavior data, explicit ratings and reviews, or provided directly by the user.
The key principle is for people to more easily scrutinize representations of their interests---to understand what is happening, correct the system, and enable capturing aspirations not reflected in past behavior without significant reduction in recommendation quality. The algorithmic requirement enabling this is that personalization is made solely\footnote{The value, as well as the limitations introduced by this constraint are discussed directly in Sec.~\ref{subsec:novelty}.} on the basis of natural language summaries.

\begin{table*}
\caption{Illustrative examples of traditional rating-based observations being summarized in natural language, scrutinized by a user, then transformed into a more informative descriptions yielding better recommendations in two domains.}
\label{tab:running}
\rowcolors{2}{gray!20}{white}
\footnotesize
\begin{tabular}{c|p{9.0cm}:p{5.6cm}}
\toprule
\multicolumn{1}{c}{\bf Recommendation\!\!\!} & \multicolumn{2}{c}{\bf Domain}\\
\multicolumn{1}{c}{\bf Stage} & \multicolumn{1}{c:}{Movies} & \multicolumn{1}{c}{Events}\\ 
\midrule
\begin{minipage}{2cm} \centering Item Ratings \end{minipage} & 
  Shawshank Redemption (+), Persona (+), Star Trek (+)\newline 
  The Piano (+), Legend of 1900 (+), Kill Bill (+)\newline
  The Hangover (-), Saving Private Ryan (-), Memento (-) & 
  
  Cirque du Soleil (+), The Lion King musical (+)\newline
  Monster Truck (+), Cambridge Beer Festival (+)\newline
  Arsenal vs Chelsea football (-), Stomp (-)\\

\begin{minipage}{2cm} \centering Initial Generated NL Summary \end{minipage} & 
  I like dramas involving unique personalities, science fiction, and movies about prison. I don't like romantic comedies or movies about World War II. I don't like complicated movies. &
  
  I like musicals and acrobatics. I also like beer events and stunt driving. I don't like football events nor music events.
  \\

\begin{minipage}{2cm} \centering Initial Recommendations \end{minipage}  & 
  Mission to Mars, The Waterboy, The Green Mile, Felon &

  Wicked (musical), Traveling circus, Heineken brewery tour, Wimbledon tennis\\
\begin{minipage}{2cm} \centering User-Scrutinized NL Summary \end{minipage} & 
  I like dramas involving unique personalities. 
  I don't like movies about drunken behaviour nor young people's antics, but do like more refined romantic comedies. 
  I don't generally like movies with a lot of violence, unless they have great cinematography like Quentin Tarantino. 
  I also like positive science fiction stories, tho not dystopian ones, especially at the end of the week. 
  I also like Pixar movies but have seen them all. &
  
  I like musicals and acrobatics.
  I make my own beer, so enjoy microbrewery events and tours.
  I like motor sports, racing and stunt driving,
  although I don't like any other sports events.
  I like going to outdoor markets on sunny days, but wouldn't travel far for one.
  \\
\begin{minipage}{2cm} \centering Scrutinized NL Recommendations \end{minipage} & 
Roman Holiday, No Country for Old Men, Arrival, The Iron Giant
 & Billy Elliot: The Musical, Saturday SoHo Farmer's Market, Brewery Tour with Mikkeller\\
\bottomrule
\end{tabular}
\end{table*}

\begin{figure}[t]
    \centering
    \includegraphics[width=0.45\textwidth]{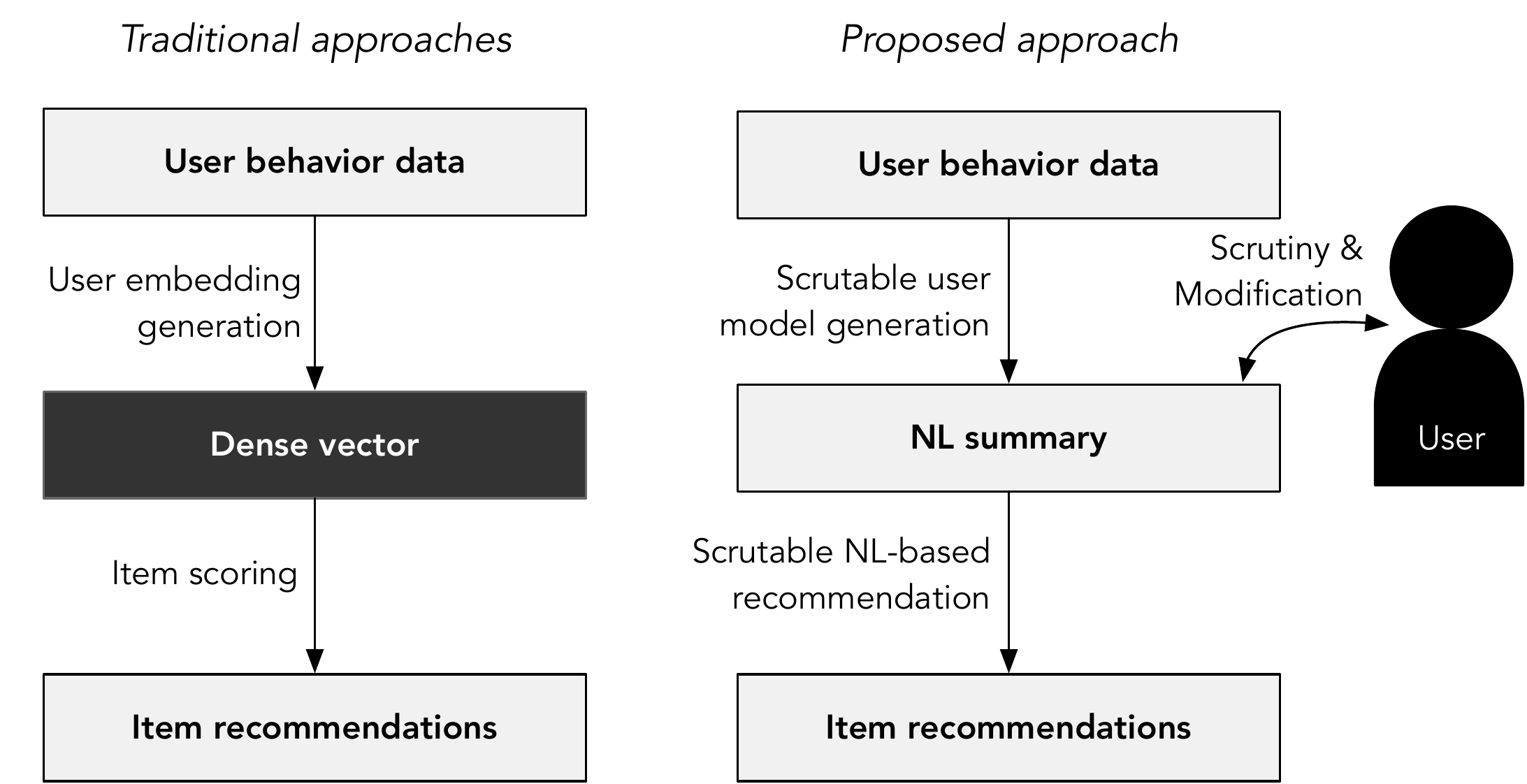}
    \caption{High-level overview of traditional approaches vs. our proposal. Instead of relying on latent representations of user preferences, we propose natural language as an explicit, transparent, and editable intermediate representation.}
    \label{fig:overview}
\end{figure}

\subsection{Main Benefits over Existing Approaches}

We now review the main benefits of the proposed approach over contemporary recommender and personalization systems, which primarily rely on user-item ratings or assumptions about implicit interaction-based indicators of user-item match~\cite{recsys-handbook}.

The most obvious benefit is \emph{transparency}: by representing knowledge about users in natural language, it can be clearer to recipients (i.e.,~users) what information is being used to generate recommendations. While explainable recommendation is a well recognized research area~\cite{zhang:explainable-recsys}, most work has focused on explaining \emph{item recommendations}, and often only captures a simplified summary of algorithm behavior as \emph{post hoc} justifications.
We focus on \emph{user-level summaries}, as first suggested in \cite{Balog:2019:SIGIR}, to suggest treating the user model in an easily accessible form: natural language. 
Such summaries would ideally make clear the aspects of a person's interests involved in the recommendation process. If successful, it would make explicit---and thus enable the elimination of---any undesirable criteria.
At the same time, natural language summaries could support highly personalized \emph{item-level explanations}, grounded in specific user model statements (see, e.g., Table~\ref{tab:lambda}).

A second benefit with more direct impact on recommendation quality is that such an approach can permit user summaries to be \emph{scrutinized} (i.e., inspected and modified), most obviously to correct errors. Expanding significantly upon \citet{Balog:2019:SIGIR}, we argue that it would be valuable for recommender systems to allow \emph{any free-form changes}. This may provide particular value for systems that rely on implicitly observed user interactions. For instance, it has been observed that a temporary interest in a topic can lead to recommender systems presenting that topic over an extended period~\cite{mcnee:hci-recsys}. If a person could easily scrutinize their model, they could eliminate such incorrect inferences. 
While keyword-based introspection has been used for control of recommendations (e.g.,~\cite{Bostandjiev:2012:RecSys}) and advertisements (e.g.,~\cite{parra:2017:myadchoices}), in practice the volume of keywords at different levels of granularity can lead to complex user interfaces that require a high amount of effort to modify downstream outputs. With this in mind, we argue that in order for the scrutability benefit to be fully realized, the effort required to inspect and modify user representations to achieve the desired result downstream must be minimal in order to motivate meaningful user adoption.

If editing preference representations improves users' recommendations, then an additional systemic benefit emerges: there is an increased incentive for people to express their preferences. This in turn can improve future recommendations.
Providing a feedback mechanism on aspects of the NL user summary could invite efficient user feedback about the user model---both about aspects that are correct and incorrect. This novel type of feedback has the potential to create yet more value for recipients of recommendations.

A related benefit of direct user control is the ability to handle \emph{significant shifts} in user preferences. It has been noted that people may find their interests change, for instance due to changing relationships, context, and so forth \cite{pereira:2018:preference-dynamics,hariri2015adapting}. 
In traditional recommendation or personalization algorithms, there are often few effective ways for someone to maintain recommendation quality while altering preferences significantly. Given a succinct user model, we hypothesize that it would be possible to modify just enough to cause the desired shift in recommendation behavior.
An additional advantage of natural language user models is that they allow for explicit expressions of aspirations (that is, allowing a distinction between what someone would like to have recommended in the future versus how they behaved in the past)~\cite{ekstrand:2016:better-recommendations}.

Finally, the existence of a user-accessible summary  could also improve \emph{cold-start recommendations}: the person may choose to describe their interests, bootstraping their recommendations. As discussed in more detail later, it may also be more useful for people to write a new, or copy an existing NL user summary.

\subsection{Research Context}

It is important to consider the question of what has changed recently to allow this fundamentally different way of representing knowledge for recommendation. 

Foremost, the ability of machine learning systems to interpret, generate, and reason about natural language text has advanced tremendously, and has manifested in systems like GPT-3~\cite{gpt3-NEURIPS2020_1457c0d6} and LaMDA~\cite{thoppilan:2022:lamda}. Much smaller models trained to look up entities have also shown promise on traditional recommendation tasks~\cite{doc-ent-2020}. This progress suggests the plausibility of recommendation systems that use natural language as the primary knowledge representation.

At the same time, transparency and user control have received significant recent interest from the research community \cite{odonovan:2005:trust,zhang:explainable-recsys}. Natural language representations of user interests offer the potential for a step change in how this can come about.

Finally, we observe that over the last few years new experimental approaches have led to datasets that, for the first time, capture both natural language and item-based knowledge about users, allowing the approach that we advocate to be evaluated directly \cite{Balog:2019:SIGIR,Balog:2020:SIGIR,Bogers:2017:narrative-driven,Radlinski:2019:CCPE}.

\subsection{Main Components}

The challenge we describe can naturally be decomposed into two components, illustrated in Figure~\ref{fig:overview}, corresponding to (1) \emph{scrutable user model generation}, and (2) \emph{scrutable NL-based recommendation}. We illustrate these tasks with running examples in Table \ref{tab:running}, showing how someone could be provided with a NL summary of their interests, modify it, and subsequently have a better experience.\footnote{While Table~\ref{tab:running} considers two popular recommendation domains, we argue that such an approach can be appropriate for almost any domain where user interests or needs can be summarized in natural language.}

\begin{definition}[Scrutable user model generation]
Given a set of interactions between a user and items (e.g., ratings, reviews, view-time, etc.) as well as metadata about items (e.g., product descriptions), \emph{scrutable user model generation} refers to the task of generating a natural language summary (\emph{NL summary}) within a given length bound that accurately summarizes a user's preferences in support of future recommendations and user scrutiny.
\end{definition}

\noindent It is important to note that succinctness (ensured by having a maximum length) is a necessary condition of the representation. Intuitively, the longer the NL summary, the more information about a user's preferences can be accurately captured, but the increased length also increases the cognitive load needed to scrutinize---this trade-off is one of the key challenges (see Sec.~\ref{sec:challenges:trade_off}). A key underlying assumption behind our approach is that the NL summary is short enough for the user to meaningfully scrutinize it, in particular to understand and modify their representation. Similarly, the scrutinization process must be easy enough, and result in visible improvements for users quickly enough, to motivate them to put in such effort.

We note that scrutable user model generation depends on the existence of text that can characterize items and their attributes. For example, this could be learned by a model from a knowledge base, associated reviews, explicit feedback given by users, etc. The source of such text is purposefully not made part of the definition, although we elaborate on various sources in Sec.~\ref{sec:challenges:item_text}.

NL summaries may be evaluated \emph{intrinsically}, in terms of the quality of the generated text, either overall or along specific dimensions (e.g., fluency, coherence, correctness)~\citep{celikyilmaz2021evaluation}.
Ultimately, NL summaries also need \emph{extrinsic} evaluation, measuring their utility for the end-to-end task of recommending useful items.
Additionally, \emph{scrutability} must clearly be a criterion: To what extent does the summary enable someone to understand their recommendations, and to what extent does it allow them to correct or update it?

\begin{definition}[Scrutable NL-based recommendation]
Given a NL summary representing a user's preferences, \emph{scrutable NL-based recommendation} refers to the task of generating a partial ordering over a set of candidate items that can be recommended. 
\end{definition}
\noindent Typically the ordering will be a top-N ranked list of item recommendations or a scoring of the items, based on predicted utility. For instance, see the ``Recommendation'' rows in Table~\ref{tab:running} that could be generated from the preceding row's NL summary.
This task can naturally be evaluated in terms of effectiveness as a top-N ranking task, using standard retrieval measures (see Sec.~\ref{sec:task2:eval_criteria}).

\vspace{\baselineskip}
\noindent
\textbf{Joint development.} The two components, generation and recommendation, are related to well-studied problems, as discussed in Sec.\,\ref{sec:task1} and \ref{sec:task2}.  
However, our proposal is unique in that generation and recommendation jointly depend on the textual representation. Generation must produce text that is both interpretable and modifiable, and effective in producing good recommendations, and for its part the NL-based recommender can only be as good as the NL summaries it is provided with.

\section{Scrutable User Model Generation}
\label{sec:task1}

We start by considering in more detail the question of generating a natural language user summary, such as those illustrated in Table~\ref{tab:running}. Specifically, we propose the research problem of the automated generation, and updating of, scrutable natural language user summaries. 
For a given user, the output is a proposed new, or updated, natural language representation that they can scrutinize.

\subsection{Research Context}

Producing a description that is correct yet easy to interpret, captures someone's breadth of interests, and can be leveraged by a recommender to produce near state of the art recommendations is an open problem. To train a model that produces such a representation one would expect to require significant training data, which may appear difficult to obtain. However, we note three branches of work that give reason to be optimistic, and we argue that these may productively be brought together to tackle the open problem.

\subsubsection{Fluent Generation: From Templates to Large Language Models}
\citet{Balog:2019:SIGIR} recently proposed the notion of transparent, scrutable, and explainable user models for recommendation. Although they used a restricted representation of users as weighted pairs of tag interactions, the authors showed how representations can be verbalized as NL statements using templates (e.g., ``You don't like [tag1] unless [tag2], for example [movie]'').
Editing was limited to removing parts of these statements, which allowed the recommendation model to still use only (pairwise) tags.\footnote{It should also be noted that the recommender algorithm uses tag weights from the underlying user model, relying on information not contained in the textual summary.} 

A key difference with the vision we lay out is that we are proposing that the generation of user summaries could be powered by advances in natural language generation rather than being template-based. This, we argue, can obtain much more nuanced statements of preference, allowing much more precise characterization of user interests. We will consider how these can be used for recommendation in Sec.~\ref{sec:task2}. In particular, we note that recent large pre-trained language models have been shown to have impressive generative fluency~\cite{ribeiro:2020:investigating, thoppilan:2022:lamda} as well as effective few-shot learning capabilities~\cite{gpt3-NEURIPS2020_1457c0d6}.
There has also been a recent branch of work that shows that a tiny fraction (e.g., 0.01\%) of the model's parameters can be tuned to provide comparable quality to 
full fine-tuning~\cite{prompt-tuning-lester-etal-2021-power}, 
which suggests that bridging between vector-based models and textual ones for user representations may be feasible. 
We also note that while language model advances have often focused on chat settings, recent work has also illustrated recommendation use cases~\cite{thoppilan:2022:lamda}.

\subsubsection{Effective User Characterization}

It has long been recognized that tags can be used effectively for recommendation~\citep{firan:2007:tag-based,kim:2009:triptip,durao:2009:personalized,Bogers:2018:Tag-Based}.  They can also be used for explaining recommendations---as particularly well studied in the movie domain~\citep{Gedikli:2014:IEC,Vig:2009:tagsplanations}---as well as for tasks such as assisting user decision making~\cite{Alhindi:2015:Profile}.
Early work on preference learning suggests that it is easier for users to express preferences over item sets than on individual items~\citep{desJardins:2006:LUP,Brafman:2006:POS}.
Indeed, \citet{Chang:2015:UGI} demonstrate that new users can complete preference elicitation more effectively by expressing preferences for groups of items (tags) rather than rating individual items.
Given that tags are a shared expression of social communities, the tags users self-assign are highly influenced by the community~\citep{Sen:2006:CSCW,Wagner:2014:WWW}. %
\citet{Sen:2007:GROUP} argue that systems should also support negative user ratings on tags, not only positive ones.
By using NL summaries, users are not limited to existing tag vocabularies and can also express the intensity of their preferences (e.g., `don't like' vs. `hate').

\subsubsection{User Control}

Past research shows that providing users with more control leads to higher  satisfaction~\citep{Knijnenburg:2012:RecSys,Bostandjiev:2012:RecSys,Dooms:2014:RecSys,Harper:2015:RecSys,Parra:2015:IJHCS}.
\citet{Jannach:2017:EWT} define user control mechanisms as requiring to have an \emph{immediate effect} on the recommendations.  
In many real-world recommender systems, users have no other means to control the behavior of the system than to change their past ratings---this does not count as a control mechanism ``as usually these changes are not immediately reflected in the results''~\citep{Jannach:2017:EWT}.
User control may be exercised during the preference elicitation phase, where users can explicitly declare their preferences on certain content categories or assign weight to item attributes~\citep{Hijikata:2012:SAC,Jannach:2017:EWT}.
Another form of control is to dynamically adjust the recommendations once they have been presented, using various UI elements for fine-tuning results~\citep{Bostandjiev:2012:RecSys,Dooms:2014:RecSys,Harper:2015:RecSys,Jin:RecSys:2018}.
\citet{Harambam:2019:RecSys} experiment with distinct control mechanisms for different phases in the recommendation process: input, recommender algorithms, and output.
While having control empowers users, having too many controls also increases cognitive load~\citep{Jin:RecSys:2018} and the design of intuitive yet easy-to-use interfaces remains an open challenge and an active research area~\citep{Jannach:2017:EWT,Lukoff:2021:CHI}. Our proposal provides control through scrutable NL summaries, allowing users to make updates and immediately see the results.

\subsection{Evaluation and Datasets}
\label{sec:task1:eval}

\subsubsection{Evaluating NL Summaries}
Scrutable user representations play two key but very different roles: on the one hand, they support good recommendations; on the other, they should be scrutable, supporting the user's understanding and control.

A key evaluation of the \textit{effectiveness} of generated NL summaries is \emph{extrinsic}, that is, via the quality of the recommendations it allows.  Because the generator's input includes interactions (e.g., ratings), recommendation quality is likely to depend on the number of past interactions.  We thus propose to evaluate effectiveness as a function of past interactions, which also allows for a direct comparison with contemporary (rating-based) recommender systems.  It is expected that, while the performance of contemporary recommender systems typically increases with the number of past interactions, it may be increasingly challenging to effectively capture additional information in a NL summary. This, however, remains an open question that can only be answered empirically.

\textit{Scrutability} of a NL summary involves assessing its ability to be both understandable and editable. 
A better ground truth than asking users to write preferences may be for them to select or edit pre-written phrases: while domain experts (e.g., movie critics) can fluently express nuanced characterizations, users may struggle with precise vocabulary to describe their tastes beyond general terms, e.g., ``I like sci-fi'' \cite{Radlinski:2019:CCPE}. 
It may be possible for such phrases to be extracted from reviews (e.g., by methods like~\citep{Narang2020WT5TT}), be written by experts, or even be directly generated by specialized language models (cf.~Sec.~\ref{sec:feasibility}). This turns the significant challenge of writing preferences into one of selecting and validating them. Editability aims to provide users with control over their recommendations, by altering their profiles (i.e., going from Row 2 to Row 4 in Table~\ref{tab:running}).  
That said, the ambiguity of language is also important to consider~\cite{Radlinski:2022:Subjectivity}, to ensure that the users and the system interpret the meaning of language consistently. Otherwise, a user's expression of a preference may lead to poor recommendations---suggesting that correctness of interpretation must also be evaluated as part of scrutability. 

Scrutability can be evaluated quantitatively, by measuring how effective an edit was in achieving the desired outcome---e.g., by removing some specific unwanted items from the suggestions, or improving the quality of recommendations as a whole, as in~\citep{Balog:2019:SIGIR}.  The usability and effectiveness of the user interface also plays an important role here, and would require an appropriate evaluation. This is key, as both the effort involved and how quickly users see a result from scrutinizing their summary are likely to strongly influence whether or not users actually scrutinize the model, even if it is possible through natural language.

\subsubsection{Evaluating Explanations}
While receiving good recommendations is key, transparency and trustworthiness are also key user considerations~\citep{Sinha:2002:CHI,Ngo:2020:UMAP}.
This is where explanations can help users to understand why a given item was suggested~\citep{Tintarev:2015:ERD}.  The NL summary establishes a shared language between the user and the system, which in turn allows for a greater degree of personalization to the individual.
Item-level explanations have been evaluated in the past along different dimensions, such as transparency, scrutability, or effectiveness~\citep{Tintarev:2015:ERD}, as well as on how personalized they feel~\citep{Tintarev:2012:UMUAP,Balog:2020:SIGIR}---we argue that similar evaluation criteria should be adapted for user-level summaries.

\subsubsection{Existing Datasets}

Generating high quality datasets of natural language representations is challenging. However, some existing datasets already highlight preliminary evaluation possibilities.

\citet{Radlinski:2019:CCPE} used a Wizard-of-Oz approach to mimic a digital assistant that elicits movie preferences.  While this study does not yield NL summaries, it provides useful insights into how users describe preferences in the movie domain. Separately, \citet{Balog:2019:SIGIR} generate template-based NL user summaries from item ratings, with users asked to scrutinize their summaries by inspecting individual sentences, and selecting a variant that most accurately described their preferences. Then, the corrected summaries are compared against initial ones in terms of recommendation quality. Narrative-driven recommendation datasets could also be used, trying to infer summaries from recommendations that follow~\citep{Bogers:2017:narrative-driven,Afzali:2021:SIGIR}.

\subsubsection{Limitations and practical considerations}

\citet{ben-akiva:sp-rp-ranking} note that stated preferences (what people say they prefer) may be inconsistent with their revealed preferences (what they actually select). This highlights a challenge with the proposed approach: people may act inconsistently. %
The effect is twofold: on the one hand, NL summaries that accurately reflect user behavior may not match what a user would say they like. Conversely, if the NL summary does accurately reflect a user's expressed interest, what they choose to consume may not match their initial expressed preferences. A related problem is that by leveraging language models, additional concerns about the unintended biases or other possible harms of utilizing such systems may need investigation  \cite{zhao-etal-2018-gender,bender:stochastic-parrots}. For instance, when characterizations are associated with specific items, it would be important to ensure that those characterizations are correct.

\section{Scrutable NL-based Recommendation}
\label{sec:task2}

The second component, \emph{scrutable NL-based recommendation}, is concerned with  generating novel item recommendations based on a natural language user preference summary, reflected in moving from the second to third and fourth to fifth rows of Table \ref{tab:running}. It involves a deep understanding of natural language and how to interpret it.  We discuss similarities and key differences to related work, then present evaluation objectives and related datasets.

\subsection{Research Context}
\label{sec:task2:related}
Although natural language is a novel way to encode long-term user preferences, matching text against items has many connections to existing work. We review these similarities and contrast scrutable NL-based recommendation.

\subsubsection{Filtering}
Originally proposed by \citet{luhn:SDI}, information filtering refers to detecting documents relevant to an information need. %
In early systems, user profiles were constructed %
by selecting keywords in context, which were then matched against incoming documents~\cite{hensley:sdi-review}.  Some years later, the Text REtrieval Conference (TREC) included a document routing task, bringing a common evaluation protocol, data, and metrics to information filtering~\cite{voorhees:trecbook}. Filtering profiles included %
keyword-like representations as well as longer sentence or paragraph-length profiles about user needs~\cite{harman:trec-1-data-description}.  Underlining the similarity between filtering and recommendation, these topics were also used for TREC \textit{ad hoc} search tasks.  Here longer, user-composed `narrative' descriptions included a variety of language, including restrictions, negations, and other modifiers~\cite{harman:trec-1-data-description}.  When provided the opportunity to amend profiles, early information filtering system users were not able to effectively improve performance~\citep{hensley:sdi-review}. More generally, later studies have shown that personalization systems that rely on content-based models can improve search results~\cite{Teevan:2005:Personalizing,Matthijs:2011:Personalizing}.

Although work in profile-based information filtering has similarities to scrutable NL-based recommendation, there are substantial differences.  To start, we do not necessarily constrain our work to text documents or even streams of items.  This means that, while early filtering systems may be similar in spirit, the precise techniques will be quite different from those specific to text.  More importantly, we believe that language technology has sufficiently advanced that even early filtering approaches deserve reevaluation.  This is especially true of previous results around user-editing of profiles, which may have suffered from keyword-based profiles.

\subsubsection{Verbose Query Retrieval}
Given the similarity of recommendation and search \cite{Belkin:1992:ACM,Furner:2002:JASIST,Bogers:2017:narrative-driven,Zamani:2022:FnTIR} and the interchangeable nature of filtering profiles and queries, we next examine relevant work in information retrieval. %
Retrieval with verbose natural language queries has attracted much attention~\citep{Gupta:2015:FnTIR},
yet the definition of a verbose query differs across retrieval contexts.  In web search, queries with five words are considered long~\citep{Gupta:2015:FnTIR}.  Narrative queries in the TREC Robust track were about 40 words long \citep{ROBUST2004}.  In clinical retrieval, queries as long as 60 words occur~\citep{kuhn:negative-feedback}, while in community QA the median question length may be over 150 words~\citep{Trienes:2019:IUQ}.

There are studies showing that verbose queries perform better than short queries, e.g.,~\citep{Zhai:2004:TOIS}.
However, in an analysis of hundreds of queries in TREC datasets, \citet{Bendersky:2008:SIGIR} find that short queries, using the `title' of the TREC topic definition, consistently outperform long `description' queries.
Another study on web search query logs shows that search engine performance drops significantly with increased query length~\citep{Bendersky:2009:WSCD}.
Other techniques developed specifically for verbose queries include query reduction (e.g., deleting terms~\citep{Yang:2014:ECIR} or selecting key noun phases~\citep{Bendersky:2008:SIGIR}) and assigning different weights to query terms based on their estimated importance~\citep{Paik:2014:CIKM,Bendersky:2010:WSDM,Zhao:2010:CIKM}.  In general the brittleness of retrieval systems to verbose queries may be due to the presence of more complex search strategies such as negation and hedging \cite{arguello:extratopicality}. %
Specifically, while verbose queries, such as found in TREC, can include `negative' preferences, retrieval performance can drop substantially when these expressions are included for standard retrieval engines~\citep{kuhn:negative-feedback}.  %

While verbose query retrieval is related to scrutable NL-based recommendation, there are substantial differences.  On the one hand, as with filtering, verbose query retrieval emphasizes text document retrieval, while our framing is much more general, including a variety of media.  More specifically, information retrieval usually focuses on acute, focused information needs.  Acute or ephemeral information needs contrast with the more persistent standing preferences found in recommendation.  Focused information needs are expressed as conjunctions of concepts or facets with a small set of relevant documents.  In our setting, a natural language profile covers a broad disjunction of preferences with a set of overlapping clusters of relevant items.  This property of user profiles captures the importance of diversity and serendipity in recommendation~\citep{li:2019:serendipity,Parapar:2021:DiversePreferences}.

\subsubsection{Narrative-driven Recommendations}

The task of \emph{narrative-driven recommendations} (NDR), introduced by \citet{Bogers:2017:narrative-driven}, describes a ``scenario where the recommendation process is driven by both a log of the user’s past transactions as well as a narrative description of their current needs or interests.'' %
NDR may be regarded as a specific form of context-aware recommendations, where recommendations not only correspond to a user's (implicit) preference profile (``items I would like''), but are also tailored to a given situation or context.
NDR could also be seen as a personalized search task, given that the user is explicitly soliciting suggestions.

Common to NDR and scrutable NL-based recommendation is the recognition of the power of natural language in describing preferences and the context for the desired recommendations.
A key difference, however, is that NDR maintains transaction logs as one of the main data sources (in addition to narrative descriptions). In scrutable NL-based recommendation we rely solely on the NL summary as input to the recommendations.
Narrative-driven recommendations are most commonly solicited on community QA fora~\citep{Bogers:2017:narrative-driven,Afzali:2021:SIGIR}, where there is often back-and-forth between the requester and those who give recommendations. We envisage that such user feedback could be elicited using conversational interfaces and in turn be utilized to update the NL summary (cf. Sec.~\ref{sec:challenges:updating}).

Similarly, our proposal contrasts with \emph{conversational recommendation}~(e.g.,~\cite{Sun:2018:Conversational}), where the user's interests may be solicited in a natural language back and forth. However, such systems tend not to summarize the need in natural language in a way that can be scrutinized and freely updated before results are selected for users.

\subsubsection{Large Language Models}
\label{sec:task2:llms}

Recent advances in language modeling provide qualitative examples that suggest large language models can recommend items based on short utterances and dialog~\cite{thoppilan:2022:lamda}. Our work proposes a significant extension of this hypothesis: that such models could allow NL summaries to become a shared scrutable object for both people and recommenders.

\citet{sachdeva2020usefulReview} provide a review of approaches leveraging review text to develop user summary highlights and conclude that the progress to date has been relatively small. The position we take, however, is \emph{not} that recommendations would necessarily need to be significantly \emph{better}, but that a NL-based user model may offer both \emph{good} recommendations, and \emph{scrutable} ones, allowing users to inspect and edit their representation. 

The closest related work that uses natural language to generate recommendations is that of \citet{doc-ent-2020}. They use a BERT-style language models to generate movie recommendations from free-form descriptive text. The main differences to our case
is that these descriptions should holistically capture the users' broad interests.
\citet{Narang2020WT5TT} demonstrate that large language models can also successfully extract relevant sub-parts of text (e.g., sentiment of movie reviews). Although they did not apply their models to extracting preferences, doing so would provide a significantly cleaner dataset for approaches like \cite{doc-ent-2020}. Moreover, this work combined with recent parameter efficient training methods, like that of \citet{prompt-tuning-lester-etal-2021-power}, suggest that significant further quality improvements are attainable.

\subsection{Evaluation and Datasets}
\label{sec:task2:eval}

We define scrutable NL-based recommendation as a ranking task, with a system presenting items in decreasing order of predicted utility.  In that sense, we can evaluate the effectiveness of a system ranking using standard information retrieval metrics such as reciprocal rank %
and nDCG. Yet we now explore how moving from relevance to utility requires a shift in evaluation methodology.

\subsubsection{Evaluation Criteria}
\label{sec:task2:eval_criteria}

Modern systems commonly model how users interact with system rankings, often adopting retrieval metrics (such as precision or nDCG)~\citep{Valcarce:2018:RecSys}.  Although an item rating, like relevance, encodes the affinity of a user to an item, when seeking recommendations, users are usually explicitly looking for items they have \textit{not} previously experienced \cite{garcia-gathright:sigir2018,leqi:satiation}.  This means that, unlike in search, the utility %
from an item depends on whether or how frequently an item has been experienced by the user.  
Additionally, the diversity of the recommendations may also be considered~\citep{Castells:2015:RecSysbook}.

\subsubsection{Existing Datasets}

Several existing datasets can be leveraged for evaluating scrutable NL-based recommendation.  As mentioned above, TREC filtering and \textit{ad hoc} retrieval datasets include verbose narratives that contain challenging language, including restrictions and difficult to support
negations~\cite{harman:trec-1-data-description}.  These collections, including the TIPSTER data~\cite{tipster},  deserve revisiting with modern language modeling techniques.  While the relevance criteria are different, these collections are still useful from a language understanding perspective.
Particularly relevant are narrative-driven recommendation datasets for %
books~\citep{Bogers:2017:narrative-driven} and points-of-interest~\citep{Afzali:2021:SIGIR}, as the ground truth aligns with our definition of item utility.  The difference, and thus limitation of these, lies in the nature of the narratives, which focus on ephemeral contextual information needs as opposed to standing preferences.
Closest to our setting, \citet{Balog:2020:SIGIR} solicited self-provided descriptions of liked and disliked items (movies) from users, along with specific item examples.  This somewhat simplified version of our task %
also had recommendations made by humans from a small set of (60) candidate items.  %
Nevertheless, the methodology can be adapted to facilitate dataset construction on a larger scale.
Note that all the above are offline test collections, which focus on this task in isolation.  We discuss desiderata for more holistic evaluation in Sec.~\ref{sec:challenges:evaluation}.

\section{Challenges and Opportunities}
\label{sec:challenges}

In presenting scrutable recommendation, we touched upon a number of challenges. We return to some of these, suggesting a research agenda toward realizing the benefits of scrutable recommendation.

\subsection{Generating Natural Language Summaries}

\subsubsection{Characteristics of NL Summaries}
\label{sec:challenges:item_text}
The first significant challenge is what type of terminology is optimal in the natural language summary? Past work has focused on attributes and social tags \cite{Bogers:2018:Tag-Based,Balog:2019:SIGIR}, but this may not provide sufficient granularity or expressiveness of preferences. While text reviews, search queries, and crowdsourced critiques are closer to a narrative description, they usually do not express preferences with as much detail as an expert in a given domain might when summarizing a particular taste refinement. Transcripts of user-generated descriptions of preferences collected by \citet{Radlinski:2019:CCPE} suggest a particular lack of formality and specificity in many user-generated descriptions.
The same corpus illustrates that most people rarely referred to metadata frequently used to characterize movies in expert-written text, such as actor and director names. This raises the question of what \emph{sort} of language would be best for summaries of user preferences? And, how can these summaries be made fluent, precise, and using the terminology that people would choose?
Similarly, could meta-preferences---such as preferences for novelty or tolerance of repetition (say when listening to music)---be captured?

\subsubsection{Cold-start Recommendations}
New users provide a particularly impactful case for NL summaries. In cold start situations, users would, by definition, lack substantial input to provide to a recommendation algorithm. This begs the question of how to solicit useful information from them. The work by \citet{Bogers:2017:narrative-driven} suggests that NL preference communication can require many requests, with clarification and refinement. How an algorithm would efficiently solicit preferences is unclear:
Should a system even solicit, or would it be better for new users to simply write text? Past Wizard-of-Oz studies have first solicited examples, then explanations of these \cite{Radlinski:2019:CCPE}. Recent work has also generated user summaries using additional information (e.g., age and occupation) along with a few reviews, or user information from auxillary domains \cite{lee2019melu, 10.1145/1352793.1352837}. %

In contrast to the user cold-start setting, other challenges come up in characterizing new items: if a user indicates a preference for/against a particular item, a system would need to know how to generalize to other items. If a new item has few existing reviews or other details available, this task would be particularly challenging.

\subsubsection{Cognitive Load}
An overarching assumption has been that text is desirable, and mathematical representations are not. It assumes text is more scrutable and requires less cognitive effort to improve and control recommendations. A key question %
is whether it really has sufficiently low cognitive load. %
Indeed, while users may want to understand and control their user summary, approaches that require large effort are likely undesirable. Therefore, it is important to consider ways to reduce the cognitive load required to generate, understand, and update user summaries. One obvious way is to ensure that representations are relatively concise. Other methods may include consistent phrasing, %
or providing multiple phrasings so that users can pick those that appeal to them. 
Other attributes of text such as readability, complexity of language, linguistic forms, and choice of terminology may also influence cognitive load, and how to trade off these aspects with the informativeness of a summary may need to be addressed.

Indeed, just text may \emph{not} be optimal: \emph{multi-modal representations} may be desirable, such as text combined with other user-defined items that represent interests---e.g., images or even audio snippets in a music recommendation setting. Any representation that maintains scrutability but improves performance or reduces cognitive load is likely valuable.%

\subsection{Updating Natural Language Summaries}
\label{sec:challenges:updating}

A special case of generating natural language summaries is algorithmically updating existing summaries given new evidence.

\subsubsection{Managing Updates of Representation}
\label{sec:challenges:managing_updates}
A NL summary needs to adapt as user preferences change.  This means the system must recognize changes in interests, %
and either automatically adjust the profile or request confirmation from the user.
One possible avenue would be to maintain two parts of a profile: preferences directly expressed or confirmed by the user, as well as those inferred but not yet validated. Recommendations and explanations can then provide as a basis specific aspects of each, %
allowing for freshness, transparency, as well as control.

\subsubsection{Recognizing New Contexts}
A seemingly straightforward case would be if %
interactions are observed with items that do not match a profile at all. For instance, unexpected interactions with items tailored for young children. A question arises of when, and how, such preferences should be incorporated into a user model. Factoring this as a new context is likely beneficial, especially as %
many preferences are contextual (e.g.,~\cite{Stefanidis:2011:contextual-preferences}). Such an effect was observed in the travel domain for point-of-interest recommendation in a new location~\cite{sun:2020:where}. However, while a new location might suggest a short-term context ``when in Hawaii,'' a sequence of such observations over a longer time may suggest the user profile should actually capture a new context ``when on vacation.'' Hence aggregating preferences that seemingly fall into different contexts is a further challenge.

\subsubsection{Refining User Summaries}
A second situation where a user profile is likely to evolve is when expertise changes. For instance, as someone explores music in a new genre, they may refine their taste to a sub-genre, a specific era, specific artists, and so forth. As such, they may choose to represent their preferences in generic language initially, refining it over time. Suggested refinements from a generative system are likely to be valuable here, and it is an open question as to how such refinements should be derived and represented so as to capture a person's interests most effectively.

\subsubsection{Realizing Dislikes}
A final aspect of updating NL summaries relates to the value of explicit \emph{negative} preferences, as differentiated from a lack of preferences about items that a user may not know about. Negative user preferences discussed in \cite{Sen:2007:GROUP} were modeled by~\citet{Balog:2019:SIGIR}, and it is natural to ask how these trade off with more details about positive preferences. %
The trade-off with scrutability is particularly pertinent if the role of the recommender is to present broad recommendations, with users valuing serendipity and diversity~\cite{zhang:2012:auralist,parapar:2021:diversity}. %

\subsection{Utilizing Natural Language Summaries}
We move on to taking advantage of NL summaries for recommendation, expanding on Sec.~\ref{sec:task2} to focus on more open-ended challenges.

\subsubsection{Interpreting Language Robustly}
\label{subsec:robust-interpretation}
When generating recommendations from NL summaries it is important that a system reliably interprets nuanced concepts. While large language models have made tremendous strides in language understanding, %
ensuring they reliably map from terms to concepts is critical. Approaches like universal sentence encoders~\cite{cer:2018:universal} provide one way to map from text to vector input suitable for traditional recommender algorithms, but text understanding must also be robust to slight variations in wording, verifying that small changes in wording do not affect recommendations significantly, as this would limit the usefulness of users scrutinizing them. %
Noting that descriptions may be ambiguous (e.g., ``Star Wars,'' may be a movie, or a series of movies, or a fictional universe with different types of content), %
errors in understanding preferences when there is little text for any given individual have a large effect.
A specific challenge is that of subjectivity, where the same word means different things to different people. When such language is used, even a user-inspected description %
may be misinterpreted~\cite{Radlinski:2022:Subjectivity}. Recognizing \emph{when} there is potential for misunderstanding, \emph{resolving} it and even \emph{expressing assumptions} about interpretations explicitly are ways this could be addressed.

\subsubsection{Explaining Recommendations}
\label{subsec:explanations}
Recently \citet{lyu2021workflow} %
showed that human recommenders spend the most significant part of their time (38\%) explaining recommendations, suggesting these serve a critical role in the recommendation process.
While past work showed that useful explanations can be generated from non-langu\-age-based models \cite{zhang:explainable-recsys}, 
NL models that have been \emph{validated by the user} offer an even stronger opportunity:
it may be possible to explain recommendations using the \emph{user's} preferred terminology and phrasing. For instance, describing a movie as \emph{funny} may be desirable, yet different people may mean different things by this term~\citep{Balog:2021:SIGIR}. Using the user's specific meaning in explanations may be beneficial.  Similarly, it may be possible to personalize to the user's particular language preferences: some people may prefer more or less formal explanations, longer or shorter explanations, explanations that bring attention to different aspects of items, and so forth. %

Explanations can also offer recourse if a match does not exist from the user's perspective but exists from the system's perspective. For instance, an explanation may bring attention to an element of someone's profile that is not specific enough to exclude non-relevant results. As yet, approaches to enable such interaction with explanations have received limited attention (e.g., \cite{rago:2021:argumentative}).

\subsubsection{Practical Utility}
\label{subsec:novelty}
Commonly, recommendation approaches are evaluated based on accuracy~\cite{recsys-handbook}. However, as noted above, \emph{novel} items often provide most value to users~\cite{li:2019:serendipity}. A natural challenge with short text summaries of user's interests is that they could not record every item the person already knows or has previously received as a recommendation: every movie watched, every song listened to, etc. To satisfy a desire for novel content, it may be valuable for a system to have some side information. The precise form this should take, and how to process this information while respecting the goals of transparency and user control is an important challenge. 

In particular, it would be important to evaluate if observational information used in a limited manner (e.g., to filter out repeated items) is preferred over it being used for recommendation.
An open challenge is the development of metrics that then allow the utility of such additional side information to be traded off against additional complexities in how users may scrutinize and modify their information. It is also likely there are different types of additional information that could be consumed, and that could impact results in different ways, so as to combine the strengths of different recommendation approaches most effectively.

\subsection{Optimization}
For generating natural language summaries, and using them for recommendation, optimization is an important challenge.%

\subsubsection{Trading off Competing Goals}
\label{sec:challenges:trade_off}

As mentioned earlier, there are many goals for the natural language descriptions (scrutable, concise, fluent, useful for recommendation) as well as for actual recommendations (accurate, novel, diverse). These competing goals need to draw on techniques that jointly optimize many metrics. In particular, the trade-off between the goals for the user summary and those for recommendations are unclear. We hypothesize that different people may have different priorities for recommendation quality versus scrutability, suggesting that a single learned model should permit adjusting this on a per-person basis.

\subsubsection{End-to-end Optimization}
Given a suitable combined loss, the next question is how to optimize it. While we have described a two part process for recommendation, recent work on $\beta$ variational autoencoders has shown that partitioning may not be necessary: it may be possible to train a recommendation algorithm for both high accuracy and desirable properties of the user encoding (such as controllability) in an end-to-end manner. E.g., \citet{Nema:2021:Disentangling} showed how to train a $\beta$-VAE so that particular attributes can be explicitly controlled in recommendations. It is possible that a recommendation model could similarly learn a NL user profile end-to-end, encoding desirable properties in the optimization loss.

\subsection{Evaluation}
\label{sec:challenges:evaluation}

As a final set of challenges, we turn to \emph{end-to-end} evaluation (com\-ponent-level evaluation being discussed in Sec.~\ref{sec:task1:eval} and~\ref{sec:task2:eval}). %

\subsubsection{Task-Specific Time/Data Efficiency}

When a user engages with a NL summary---be it creating or modifying one---it would be important to understand the concrete utility to the user. For example, a NL summary of a given length may take a given amount of time and effort to produce. How does this compare to the number of traditional %
ratings that a person could produce in the same amount of time? Is there a relationship that a given number of language tokens is equivalent to a specific number ratings, in terms of recommendation performance? 

Notably, when there is a shift in user preferences, the effect of adding or removing a sentence from the NL summary could be contrasted against the number of ratings the user would need to change for the same effect. Perhaps recommendations from a natural language model may be poorer than state-of-the-art rating-based recommender systems in a non-cold start scenario, yet the ease of curation may close this gap for users who choose to improve their model in language rather than by changing ratings.

\subsubsection{Coverage versus Specificity}
Suppose a description could either perfectly describe a fraction of the user's interests, or cover all their preferences but less well. An evaluation must ask: Which is better? Top-N recommendations typically prefer the former, but it is more difficult to quantify the latter: What fraction of all the user's interests are represented in the output of a recommendation algorithm? %

\subsubsection{Types of Utility and Cognitive Effort}

Overall, it is important to observe that there are two types of utility: recommendation quality and language utility. From a system perspective, specific types of description elements may have the strongest effect on recommendations (say, examples of attributes that do or do not appeal to the user). From a user perspective, this may be different---it may be unnatural or difficult for someone to provide text with the right sort of language. On the other hand, other types of natural language input may be less useful to the system but easier for a person to generate or edit (say, free-form descriptions of items that appeal and do not appeal).
As such, evaluation methodologies that incorporate cognitive effort as well as time to improve are important to develop. As discussed in Sec.~\ref{sec:challenges:updating}, there may be approaches that assist in this curation process such as suggesting phrases, flagging ambiguity, or asking explicit questions.%

Similarly, as scrutable NL-based recommendations permit users to edit their own preference summaries directly, these should have a direct and immediate impact on the recommendations that they receive. Possible edits could be simulated to measure how different types of edits affect recommendations.

\section{Feasibility Study}
\label{sec:feasibility}

\begin{table}
\caption{Few-shot prompt template used to turn a large language model into a recommender.}
\label{tab:descriptions}
\small
\begin{tabular}{|  c  |}
\hline
\begin{minipage}{8cm}
\vspace{1mm}
{\sf 
"contemplative, trippy, mind-bending or artsy sci-fi movies": \\
"2001: A Space Odyssey" - a classic sci-fi movie by Stanley Kubrick, with lots of trippy and artsy moments. \\
"Primer" - a mind-bending time travel movie where timelines constantly intersect and overlap. \\
"Arrival" - a contemplative sci-fi movie about first contact. \\ 
\\
"cool sci-fi settings that are somewhat dystopic":  \\
"Blade Runner" - a film noir sci-fi, where rogue AIs are hunted down.  \\
"District 9" - aliens visit Earth, and are treated as second-class citizens.  \\
"Gattaca" - a near-future dystopia about genetics determining your fate. \\ 
\\
"funny and non-traditional superhero movies":  \\
"The Lego Movie" - basically a superhero movie, full of humor and doesn't take itself too seriously. \\
"Thor: Ragnarok" - the silliest of all the Thor movies, with some great secondary characters.  \\
"Deadpool" - the humor is over the top and obscene, with a relentlessly funny narrator. \\ 
\\
"\textbf{\emph{<NL-SUMMARY>}}":} \vspace{1mm}
\end{minipage} \\
\hline
\end{tabular}
\end{table}

We now present a small case study highlighting how a large language model can be used for movie recommendation based on NL summaries. In particular, we use LaMDA, a decoder-only transfor\-mer-based, 137B parameter language model~\cite{thoppilan:2022:lamda}. It is trained only to predict the next token in a large generic text corpus. The results from this model are similar to other large language models such as GPT-3~\cite{gpt3-NEURIPS2020_1457c0d6}. 

The model is adapted for recommendation by configuring it using ``few-shot'' prompt templates \cite{gpt3-NEURIPS2020_1457c0d6}: Table \ref{tab:descriptions} presents the exact input given to it, where the placeholder {\tt <NL-SUMMARY>} was replaced by the user's NL summary. The model is then simply used to predict the next token, repeatedly within a top-K decoding strategy. %
The output from a single run for three example NL summaries is shown in Table~\ref{tab:lambda}, showing the first four recommendations for each.\footnote{We note that while the few-shot templates were crafted, the NL summaries and outputs were not cherry-picked.}

Given that no model fine-tuning has been applied, and the model was primed with only three examples of NL summaries paired with three recommendations and brief explanations, it is impressive that it gives both appropriate recommendations and personalized explanations. For example, referencing ``creative fight scenes'' in animated movies, instead of their animation style when the user expressed interest in ``creative and interesting fight scenes.'' When contrasted to entering the NL summary as queries into popular search engines, we observe search engines provide links to pages of lists, e.g., ``space exploration movies that are funny,'' capturing aspects of the NL query but much less specifically than the language model results, and lacking explanations. In some cases, e.g., ``space exploration movies that are funny,'' search engines do interpret the query as genre constraints for a specialised movie search, in which case there is a reasonable overlap with the NL recommender's suggestions, but again without explanation.

\begin{table}[t]
\caption{Example NL summary inputs with the corresponding generated output text that includes recommendations and explanations that reference the NL summary.}
\vspace{-0.5\baselineskip}
\label{tab:lambda}
\small
    \begin{tabular}{| l p{7.7cm} |}
      \hline
      \multicolumn{2}{|l|}{\textsf{\textbf{NL Rep.:} movies where the murderer is involved from the beginning}} \\
      & 
      \begin{minipage}{7.7cm}
      \vspace{1mm}
       \textsf{\textbf{"The Usual Suspects"} - a movie where the narrator is unreliable, and the murderer is involved from the beginning.} \\ 
      \textsf{\textbf{"Se7en"} - a detective story about a serial killer, where the killer is revealed from the beginning.} \\ 
      \textsf{\textbf{"Scream"} - a classic horror movie where the killer is revealed early on.} \\
      \textsf{\textbf{"Memento"} - a movie about a man with short-term memory loss, who is trying to find his wife's killer.}
      \vspace{1mm}
      \end{minipage}\\
      \hline
      \multicolumn{2}{|l|}{\textsf{\textbf{NL Rep.:} space exploration movies that are funny}}\\
      &
      \begin{minipage}{7.7cm}
      \vspace{1mm}
      \textsf{\textbf{"Guardians of the Galaxy"} - a team of intergalactic misfits, who are funny and have a lot of heart.} \\
      \textsf{\textbf{"The Martian"} - a great adaptation of the book by Andy Weir, with a great cast.} \\
      \textsf{\textbf{"Spaceballs"} - a parody of Star Wars and other sci-fi movies.} \\
      \textsf{\textbf{"The Hitchhiker's Guide to the Galaxy"} - a funny sci-fi movie that parodies space exploration.}
      \vspace{1mm}
      \end{minipage} \\
      \hline
      \multicolumn{2}{|l|}{\textsf{\textbf{NL Rep.:} animated movies with creative and interesting fight scenes}}\\
      & 
      \begin{minipage}{7.7cm}
      \vspace{1mm}
      \textsf{\textbf{"Akira"} - a classic anime movie about a boy with telekinetic powers.} \\
      \textsf{\textbf{"The Incredibles"} - a family of superheroes with some of the best fight scenes you'll see.} \\
      \textsf{\textbf{"Kung Fu Panda"} - a kung fu movie with a panda as the protagonist.}  \\
      \textsf{\textbf{"Sword of the Stranger"} - a samurai movie, with creative fight scenes and a bittersweet ending.}
      \end{minipage} \\
      \hline
    \end{tabular}
\end{table}

While Table~\ref{tab:lambda} may give the initial appearance of having solved the NL recommendation problem, there are significant limitations. Among those described in Sec.~\ref{sec:challenges}, we particularly observe:
(1) The NL summaries above are short, focusing on a single aspect, rather than holistic summaries of real people's broad interests. System performance with more realistic user descriptions matters more (cf. Sec.~\ref{sec:challenges:item_text}). 
(2) If users prefer novel recommendations, the recommendations here may not respect this (cf. Sec.~\ref{subsec:novelty}).
(3) Language models can contain hard to identify unintended biases. For example, we observe that the movies recommended by this model may skew towards popular movies in the United States, and are less accurate if queries include specific  constraints like precise release decades (cf.~Sec.~\ref{subsec:robust-interpretation}). %
(4) Language models have been observed to hallucinate. For example, the first example in Table~\ref{tab:lambda} highlights ``Se7en'' as a movie with a killer revealed in the beginning, but the killer is revealed much later in the film (cf. Sec.~\ref{subsec:explanations}).

\section{Conclusions}
\label{sec:concl}

We  argue that recent advances in natural language generation and understanding make it possible to envision a fundamentally new approach to personalized search and recommendation, one where systems' knowledge of users is encoded in natural language. In providing a process by which people can review and update their representations, it could significant improve transparency, user control, and, ultimately, quality of recommendations. This paper presents a breakdown of the end-to-end task of scrutable recommendation into more tractable steps and identifies both promising recent advances as well as open challenges for the broader research community to address.

\section*{Acknowledgments}

We would like to thank the anonymous reviewers as well as Alex Beutel for their useful feedback on this paper.

\bibliographystyle{ACM-Reference-Format}
\balance
\bibliography{references}


\begin{thebibliography}{92}


\ifx \showCODEN    \undefined \def \showCODEN     #1{\unskip}     \fi
\ifx \showDOI      \undefined \def \showDOI       #1{#1}\fi
\ifx \showISBNx    \undefined \def \showISBNx     #1{\unskip}     \fi
\ifx \showISBNxiii \undefined \def \showISBNxiii  #1{\unskip}     \fi
\ifx \showISSN     \undefined \def \showISSN      #1{\unskip}     \fi
\ifx \showLCCN     \undefined \def \showLCCN      #1{\unskip}     \fi
\ifx \shownote     \undefined \def \shownote      #1{#1}          \fi
\ifx \showarticletitle \undefined \def \showarticletitle #1{#1}   \fi
\ifx \showURL      \undefined \def \showURL       {\relax}        \fi
\providecommand\bibfield[2]{#2}
\providecommand\bibinfo[2]{#2}
\providecommand\natexlab[1]{#1}
\providecommand\showeprint[2][]{arXiv:#2}

\bibitem[Afzali et~al\mbox{.}(2021)]%
        {Afzali:2021:SIGIR}
\bibfield{author}{\bibinfo{person}{Jafar Afzali},
  \bibinfo{person}{Aleksander~Mark Drzewiecki}, {and}
  \bibinfo{person}{Krisztian Balog}.} \bibinfo{year}{2021}\natexlab{}.
\newblock \showarticletitle{{POINTREC}: {A} Test Collection for
  Narrative-Driven Point of Interest Recommendation}. In
  \bibinfo{booktitle}{\emph{Proceedings of the 44th International ACM SIGIR
  Conference on Research and Development in Information Retrieval}}
  \emph{(\bibinfo{series}{SIGIR '21})}. \bibinfo{pages}{2478--2484}.
\newblock


\bibitem[Alhindi et~al\mbox{.}(2015)]%
        {Alhindi:2015:Profile}
\bibfield{author}{\bibinfo{person}{Azhar Alhindi}, \bibinfo{person}{Udo
  Kruschwitz}, \bibinfo{person}{Chris Fox}, {and} \bibinfo{person}{M-Dyaa
  Albakour}.} \bibinfo{year}{2015}\natexlab{}.
\newblock \showarticletitle{Profile-Based Summarisation for Web Site
  Navigation}.
\newblock \bibinfo{journal}{\emph{ACM Trans. Inf. Syst.}} \bibinfo{volume}{33},
  \bibinfo{number}{1}, Article \bibinfo{articleno}{4} (\bibinfo{date}{feb}
  \bibinfo{year}{2015}), \bibinfo{numpages}{39}~pages.
\newblock
\showISSN{1046-8188}


\bibitem[Arguello et~al\mbox{.}(2018)]%
        {arguello:extratopicality}
\bibfield{author}{\bibinfo{person}{Jaime Arguello}, \bibinfo{person}{Bogeum
  Choi}, {and} \bibinfo{person}{Robert Capra}.}
  \bibinfo{year}{2018}\natexlab{}.
\newblock \showarticletitle{Factors Influencing Users' Information Requests:
  Medium, Target, and Extra-Topical Dimension}.
\newblock \bibinfo{journal}{\emph{ACM Trans. Inf. Syst.}} \bibinfo{volume}{36},
  \bibinfo{number}{4} (\bibinfo{date}{July} \bibinfo{year}{2018}),
  \bibinfo{pages}{41:1--41:37}.
\newblock


\bibitem[Balog and Radlinski(2020)]%
        {Balog:2020:SIGIR}
\bibfield{author}{\bibinfo{person}{Krisztian Balog} {and}
  \bibinfo{person}{Filip Radlinski}.} \bibinfo{year}{2020}\natexlab{}.
\newblock \showarticletitle{Measuring Recommendation Explanation Quality: The
  Conflicting Goals of Explanations}. In \bibinfo{booktitle}{\emph{Proceedings
  of the 43rd International ACM SIGIR Conference on Research and Development in
  Information Retrieval}} \emph{(\bibinfo{series}{SIGIR '20})}.
  \bibinfo{pages}{329--338}.
\newblock


\bibitem[Balog et~al\mbox{.}(2019)]%
        {Balog:2019:SIGIR}
\bibfield{author}{\bibinfo{person}{Krisztian Balog}, \bibinfo{person}{Filip
  Radlinski}, {and} \bibinfo{person}{Shushan Arakelyan}.}
  \bibinfo{year}{2019}\natexlab{}.
\newblock \showarticletitle{Transparent, Scrutable and Explainable User Models
  for Personalized Recommendation}. In \bibinfo{booktitle}{\emph{Proceedings of
  the 42nd International ACM SIGIR Conference on Research and Development in
  Information Retrieval}} \emph{(\bibinfo{series}{SIGIR '19})}.
  \bibinfo{pages}{265--274}.
\newblock


\bibitem[Balog et~al\mbox{.}(2021)]%
        {Balog:2021:SIGIR}
\bibfield{author}{\bibinfo{person}{Krisztian Balog}, \bibinfo{person}{Filip
  Radlinski}, {and} \bibinfo{person}{Alexandros Karatzoglou}.}
  \bibinfo{year}{2021}\natexlab{}.
\newblock \showarticletitle{On Interpretation and Measurement of Soft
  Attributes for Recommendation}. In \bibinfo{booktitle}{\emph{Proceedings of
  the 44th International ACM SIGIR Conference on Research and Development in
  Information Retrieval}} \emph{(\bibinfo{series}{SIGIR '21})}.
  \bibinfo{pages}{890--899}.
\newblock


\bibitem[Belkin and Croft(1992)]%
        {Belkin:1992:ACM}
\bibfield{author}{\bibinfo{person}{Nicholas~J. Belkin} {and}
  \bibinfo{person}{W.~Bruce Croft}.} \bibinfo{year}{1992}\natexlab{}.
\newblock \showarticletitle{Information Filtering and Information Retrieval:
  Two Sides of the Same Coin?}
\newblock \bibinfo{journal}{\emph{Commun. ACM}} \bibinfo{volume}{35},
  \bibinfo{number}{12} (\bibinfo{date}{dec} \bibinfo{year}{1992}),
  \bibinfo{pages}{29--38}.
\newblock


\bibitem[Ben-Akiva et~al\mbox{.}(1992)]%
        {ben-akiva:sp-rp-ranking}
\bibfield{author}{\bibinfo{person}{Moshe Ben-Akiva}, \bibinfo{person}{Takayuki
  Morikawa}, {and} \bibinfo{person}{Fumiaki Shiroishi}.}
  \bibinfo{year}{1992}\natexlab{}.
\newblock \showarticletitle{Analysis of the Reliability of Preference Ranking
  Data}.
\newblock \bibinfo{journal}{\emph{J. Bus. Res.}} \bibinfo{volume}{24},
  \bibinfo{number}{2} (\bibinfo{year}{1992}), \bibinfo{pages}{149--164}.
\newblock


\bibitem[Bender et~al\mbox{.}(2021)]%
        {bender:stochastic-parrots}
\bibfield{author}{\bibinfo{person}{Emily~M. Bender}, \bibinfo{person}{Timnit
  Gebru}, \bibinfo{person}{Angelina McMillan-Major}, {and}
  \bibinfo{person}{Shmargaret Shmitchell}.} \bibinfo{year}{2021}\natexlab{}.
\newblock \showarticletitle{On the Dangers of Stochastic Parrots: Can Language
  Models Be Too Big?}. In \bibinfo{booktitle}{\emph{Proceedings of the 2021 ACM
  Conference on Fairness, Accountability, and Transparency}}
  \emph{(\bibinfo{series}{FAccT '21})}. \bibinfo{pages}{610--623}.
\newblock


\bibitem[Bendersky and Croft(2008)]%
        {Bendersky:2008:SIGIR}
\bibfield{author}{\bibinfo{person}{Michael Bendersky} {and}
  \bibinfo{person}{W.~Bruce Croft}.} \bibinfo{year}{2008}\natexlab{}.
\newblock \showarticletitle{Discovering Key Concepts in Verbose Queries}. In
  \bibinfo{booktitle}{\emph{Proceedings of the 31st International ACM SIGIR
  Conference on Research and Development in Information Retrieval}}
  \emph{(\bibinfo{series}{SIGIR '08})}. \bibinfo{pages}{491--498}.
\newblock


\bibitem[Bendersky and Croft(2009)]%
        {Bendersky:2009:WSCD}
\bibfield{author}{\bibinfo{person}{Michael Bendersky} {and}
  \bibinfo{person}{W.~Bruce Croft}.} \bibinfo{year}{2009}\natexlab{}.
\newblock \showarticletitle{Analysis of Long Queries in a Large Scale Search
  Log}. In \bibinfo{booktitle}{\emph{Proceedings of the 2009 Workshop on Web
  Search Click Data}} \emph{(\bibinfo{series}{WSCD '09})}.
  \bibinfo{pages}{8--14}.
\newblock


\bibitem[Bendersky et~al\mbox{.}(2010)]%
        {Bendersky:2010:WSDM}
\bibfield{author}{\bibinfo{person}{Michael Bendersky}, \bibinfo{person}{Donald
  Metzler}, {and} \bibinfo{person}{W.~Bruce Croft}.}
  \bibinfo{year}{2010}\natexlab{}.
\newblock \showarticletitle{Learning Concept Importance Using a Weighted
  Dependence Model}. In \bibinfo{booktitle}{\emph{Proceedings of the Third ACM
  International Conference on Web Search and Data Mining}}
  \emph{(\bibinfo{series}{WSDM '10})}. \bibinfo{pages}{31--40}.
\newblock


\bibitem[Bogers(2018)]%
        {Bogers:2018:Tag-Based}
\bibfield{author}{\bibinfo{person}{Toine Bogers}.}
  \bibinfo{year}{2018}\natexlab{}.
\newblock \bibinfo{booktitle}{\emph{Tag-Based Recommendation}}.
\newblock \bibinfo{publisher}{Springer International Publishing},
  \bibinfo{address}{Cham}, \bibinfo{pages}{441--479}.
\newblock


\bibitem[Bogers and Koolen(2017)]%
        {Bogers:2017:narrative-driven}
\bibfield{author}{\bibinfo{person}{Toine Bogers} {and} \bibinfo{person}{Marijn
  Koolen}.} \bibinfo{year}{2017}\natexlab{}.
\newblock \showarticletitle{Defining and Supporting Narrative-Driven
  Recommendation}. In \bibinfo{booktitle}{\emph{Proceedings of the Eleventh ACM
  Conference on Recommender Systems}} \emph{(\bibinfo{series}{RecSys '17})}.
  \bibinfo{pages}{238--242}.
\newblock


\bibitem[Bostandjiev et~al\mbox{.}(2012)]%
        {Bostandjiev:2012:RecSys}
\bibfield{author}{\bibinfo{person}{Svetlin Bostandjiev}, \bibinfo{person}{John
  O'Donovan}, {and} \bibinfo{person}{Tobias H\"{o}llerer}.}
  \bibinfo{year}{2012}\natexlab{}.
\newblock \showarticletitle{TasteWeights: A Visual Interactive Hybrid
  Recommender System}. In \bibinfo{booktitle}{\emph{Proceedings of the Sixth
  ACM Conference on Recommender Systems}} \emph{(\bibinfo{series}{RecSys
  '12})}. \bibinfo{pages}{35--42}.
\newblock


\bibitem[Brafman et~al\mbox{.}(2006)]%
        {Brafman:2006:POS}
\bibfield{author}{\bibinfo{person}{Ronen~I. Brafman}, \bibinfo{person}{Carmel
  Domshlak}, \bibinfo{person}{Solomon~Eyal Shimony}, {and} \bibinfo{person}{Y.
  Silver}.} \bibinfo{year}{2006}\natexlab{}.
\newblock \showarticletitle{Preferences over Sets}. In
  \bibinfo{booktitle}{\emph{Proceedings of the Twenty-First National Conference
  on Artificial Intelligence and the Eighteenth Innovative Applications of
  Artificial Intelligence Conference}}. \bibinfo{pages}{1101--1106}.
\newblock


\bibitem[Brown et~al\mbox{.}(2020)]%
        {gpt3-NEURIPS2020_1457c0d6}
\bibfield{author}{\bibinfo{person}{Tom Brown}, \bibinfo{person}{Benjamin Mann},
  \bibinfo{person}{Nick Ryder}, \bibinfo{person}{Melanie Subbiah},
  \bibinfo{person}{Jared~D Kaplan}, \bibinfo{person}{Prafulla Dhariwal},
  \bibinfo{person}{Arvind Neelakantan}, \bibinfo{person}{Pranav Shyam},
  \bibinfo{person}{Girish Sastry}, \bibinfo{person}{Amanda Askell},
  \bibinfo{person}{Sandhini Agarwal}, \bibinfo{person}{Ariel Herbert-Voss},
  \bibinfo{person}{Gretchen Krueger}, \bibinfo{person}{Tom Henighan},
  \bibinfo{person}{Rewon Child}, \bibinfo{person}{Aditya Ramesh},
  \bibinfo{person}{Daniel Ziegler}, \bibinfo{person}{Jeffrey Wu},
  \bibinfo{person}{Clemens Winter}, \bibinfo{person}{Chris Hesse},
  \bibinfo{person}{Mark Chen}, \bibinfo{person}{Eric Sigler},
  \bibinfo{person}{Mateusz Litwin}, \bibinfo{person}{Scott Gray},
  \bibinfo{person}{Benjamin Chess}, \bibinfo{person}{Jack Clark},
  \bibinfo{person}{Christopher Berner}, \bibinfo{person}{Sam McCandlish},
  \bibinfo{person}{Alec Radford}, \bibinfo{person}{Ilya Sutskever}, {and}
  \bibinfo{person}{Dario Amodei}.} \bibinfo{year}{2020}\natexlab{}.
\newblock \showarticletitle{Language Models are Few-Shot Learners}. In
  \bibinfo{booktitle}{\emph{Advances in Neural Information Processing
  Systems}}, \bibfield{editor}{\bibinfo{person}{H.~Larochelle},
  \bibinfo{person}{M.~Ranzato}, \bibinfo{person}{R.~Hadsell},
  \bibinfo{person}{M.~F. Balcan}, {and} \bibinfo{person}{H.~Lin}} (Eds.),
  Vol.~\bibinfo{volume}{33}. \bibinfo{publisher}{Curran Associates, Inc.},
  \bibinfo{pages}{1877--1901}.
\newblock


\bibitem[Castells et~al\mbox{.}(2015)]%
        {Castells:2015:RecSysbook}
\bibfield{author}{\bibinfo{person}{Pablo Castells}, \bibinfo{person}{Neil~J.
  Hurley}, {and} \bibinfo{person}{Saul Vargas}.}
  \bibinfo{year}{2015}\natexlab{}.
\newblock \showarticletitle{Novelty and Diversity in Recommender Systems}.
\newblock In \bibinfo{booktitle}{\emph{Recommender Systems Handbook}
  (\bibinfo{edition}{2nd} ed.)}, \bibfield{editor}{\bibinfo{person}{Francesco
  Ricci}, \bibinfo{person}{Lior Rokach}, {and} \bibinfo{person}{Bracha
  Shapira}} (Eds.). \bibinfo{pages}{881--918}.
\newblock


\bibitem[Celikyilmaz et~al\mbox{.}(2021)]%
        {celikyilmaz2021evaluation}
\bibfield{author}{\bibinfo{person}{Asli Celikyilmaz},
  \bibinfo{person}{Elizabeth Clark}, {and} \bibinfo{person}{Jianfeng Gao}.}
  \bibinfo{year}{2021}\natexlab{}.
\newblock \bibinfo{title}{Evaluation of Text Generation: A Survey}.
\newblock
\newblock
\showeprint[arxiv]{2006.14799}~[cs.CL]


\bibitem[Cer et~al\mbox{.}(2018)]%
        {cer:2018:universal}
\bibfield{author}{\bibinfo{person}{Daniel Cer}, \bibinfo{person}{Yinfei Yang},
  \bibinfo{person}{Sheng-yi Kong}, \bibinfo{person}{Nan Hua},
  \bibinfo{person}{Nicole Limtiaco}, \bibinfo{person}{Rhomni St.~John},
  \bibinfo{person}{Noah Constant}, \bibinfo{person}{Mario Guajardo-Cespedes},
  \bibinfo{person}{Steve Yuan}, \bibinfo{person}{Chris Tar},
  \bibinfo{person}{Brian Strope}, {and} \bibinfo{person}{Ray Kurzweil}.}
  \bibinfo{year}{2018}\natexlab{}.
\newblock \showarticletitle{Universal Sentence Encoder for English}. In
  \bibinfo{booktitle}{\emph{Proceedings of the 2018 Conference on Empirical
  Methods in Natural Language Processing: System Demonstrations}}
  \emph{(\bibinfo{series}{EMNLP '18})}. \bibinfo{pages}{169--174}.
\newblock


\bibitem[Chang et~al\mbox{.}(2015)]%
        {Chang:2015:UGI}
\bibfield{author}{\bibinfo{person}{Shuo Chang}, \bibinfo{person}{F.~Maxwell
  Harper}, {and} \bibinfo{person}{Loren Terveen}.}
  \bibinfo{year}{2015}\natexlab{}.
\newblock \showarticletitle{Using Groups of Items for Preference Elicitation in
  Recommender Systems}. In \bibinfo{booktitle}{\emph{Proceedings of the ACM
  Conference on Computer Supported Cooperative Work \& Social Computing}}
  \emph{(\bibinfo{series}{CSCW '15})}. \bibinfo{pages}{1258--1269}.
\newblock


\bibitem[Chen et~al\mbox{.}(2019)]%
        {li:2019:serendipity}
\bibfield{author}{\bibinfo{person}{Li Chen}, \bibinfo{person}{Yonghua Yang},
  \bibinfo{person}{Ningxia Wang}, \bibinfo{person}{Keping Yang}, {and}
  \bibinfo{person}{Quan Yuan}.} \bibinfo{year}{2019}\natexlab{}.
\newblock \showarticletitle{How Serendipity Improves User Satisfaction with
  Recommendations? A Large-Scale User Evaluation}. In
  \bibinfo{booktitle}{\emph{The World Wide Web Conference}}
  \emph{(\bibinfo{series}{WWW '19})}. \bibinfo{pages}{240--250}.
\newblock


\bibitem[desJardins et~al\mbox{.}(2006)]%
        {desJardins:2006:LUP}
\bibfield{author}{\bibinfo{person}{Marie desJardins}, \bibinfo{person}{Eric
  Eaton}, {and} \bibinfo{person}{Kiri~L. Wagstaff}.}
  \bibinfo{year}{2006}\natexlab{}.
\newblock \showarticletitle{Learning User Preferences for Sets of Objects}. In
  \bibinfo{booktitle}{\emph{Proceedings of the 23rd International Conference on
  Machine Learning}} \emph{(\bibinfo{series}{ICML '06})}.
  \bibinfo{pages}{273--280}.
\newblock


\bibitem[Dooms et~al\mbox{.}(2014)]%
        {Dooms:2014:RecSys}
\bibfield{author}{\bibinfo{person}{Simon Dooms}, \bibinfo{person}{Toon~De
  Pessemier}, {and} \bibinfo{person}{Luc Martens}.}
  \bibinfo{year}{2014}\natexlab{}.
\newblock \showarticletitle{Improving IMDb Movie Recommendations with
  Interactive Settings and Filters}. In \bibinfo{booktitle}{\emph{Poster
  Proceedings of the 8th ACM Conference on Recommender Systems}}
  \emph{(\bibinfo{series}{RecSys '14})}.
\newblock


\bibitem[Dur{\~{a}}o and Dolog(2009)]%
        {durao:2009:personalized}
\bibfield{author}{\bibinfo{person}{Frederico~Ara{\'{u}}jo Dur{\~{a}}o} {and}
  \bibinfo{person}{Peter Dolog}.} \bibinfo{year}{2009}\natexlab{}.
\newblock \showarticletitle{A Personalized Tag-Based Recommendation in Social
  Web Systems}. In \bibinfo{booktitle}{\emph{Proceedings of the Workshop on
  Adaptation and Personalization for Web 2.0}}.
\newblock


\bibitem[Ekstrand and Willemsen(2016)]%
        {ekstrand:2016:better-recommendations}
\bibfield{author}{\bibinfo{person}{Michael~D Ekstrand} {and}
  \bibinfo{person}{Martijn~C Willemsen}.} \bibinfo{year}{2016}\natexlab{}.
\newblock \showarticletitle{Behaviorism is Not Enough: Better Recommendations
  through Listening to Users}. In \bibinfo{booktitle}{\emph{Proceedings of the
  10th ACM conference on recommender systems}} \emph{(\bibinfo{series}{RecSys
  '16})}. \bibinfo{pages}{221--224}.
\newblock


\bibitem[Firan et~al\mbox{.}(2007)]%
        {firan:2007:tag-based}
\bibfield{author}{\bibinfo{person}{Claudiu~S Firan}, \bibinfo{person}{Wolfgang
  Nejdl}, {and} \bibinfo{person}{Raluca Paiu}.}
  \bibinfo{year}{2007}\natexlab{}.
\newblock \showarticletitle{The Benefit of Using Tag-Based Profiles}. In
  \bibinfo{booktitle}{\emph{2007 Latin American Web Conference (LA-WEB '07)}}.
  \bibinfo{pages}{32--41}.
\newblock


\bibitem[Furner(2002)]%
        {Furner:2002:JASIST}
\bibfield{author}{\bibinfo{person}{Jonathan Furner}.}
  \bibinfo{year}{2002}\natexlab{}.
\newblock \showarticletitle{On Recommending}.
\newblock \bibinfo{journal}{\emph{J. Assoc. Inf. Sci. Technol.}}
  \bibinfo{volume}{53} (\bibinfo{year}{2002}), \bibinfo{pages}{747--763}.
\newblock


\bibitem[Garcia-Gathright et~al\mbox{.}(2018)]%
        {garcia-gathright:sigir2018}
\bibfield{author}{\bibinfo{person}{Jean Garcia-Gathright},
  \bibinfo{person}{Brian {St. Thomas}}, \bibinfo{person}{Christine Hosey},
  \bibinfo{person}{Zahra Nazari}, {and} \bibinfo{person}{Fernando Diaz}.}
  \bibinfo{year}{2018}\natexlab{}.
\newblock \showarticletitle{Understanding and Evaluating User Satisfaction with
  Music Discovery}. In \bibinfo{booktitle}{\emph{Proceedings of the 41st
  International ACM SIGIR Conference on Research and Development in Information
  Retrieval}} \emph{(\bibinfo{series}{SIGIR '18})}. \bibinfo{pages}{55--64}.
\newblock


\bibitem[Gedikli et~al\mbox{.}(2014)]%
        {Gedikli:2014:IEC}
\bibfield{author}{\bibinfo{person}{Fatih Gedikli}, \bibinfo{person}{Dietmar
  Jannach}, {and} \bibinfo{person}{Mouzhi Ge}.}
  \bibinfo{year}{2014}\natexlab{}.
\newblock \showarticletitle{How Should I Explain? A Comparison of Different
  Explanation Types for Recommender Systems}.
\newblock \bibinfo{journal}{\emph{Int. J. Hum.-Comput. Stud.}}
  \bibinfo{volume}{72}, \bibinfo{number}{4} (\bibinfo{date}{April}
  \bibinfo{year}{2014}), \bibinfo{pages}{367--382}.
\newblock


\bibitem[Gupta and Bendersky(2015)]%
        {Gupta:2015:FnTIR}
\bibfield{author}{\bibinfo{person}{Manish Gupta} {and} \bibinfo{person}{Michael
  Bendersky}.} \bibinfo{year}{2015}\natexlab{}.
\newblock \showarticletitle{Information Retrieval with Verbose Queries}.
\newblock \bibinfo{journal}{\emph{Found. Trends Inf. Ret.}}
  \bibinfo{volume}{9}, \bibinfo{number}{3--4} (\bibinfo{year}{2015}),
  \bibinfo{pages}{209--354}.
\newblock


\bibitem[Harambam et~al\mbox{.}(2019)]%
        {Harambam:2019:RecSys}
\bibfield{author}{\bibinfo{person}{Jaron Harambam}, \bibinfo{person}{Dimitrios
  Bountouridis}, \bibinfo{person}{Mykola Makhortykh}, {and}
  \bibinfo{person}{Joris van Hoboken}.} \bibinfo{year}{2019}\natexlab{}.
\newblock \showarticletitle{Designing for the Better by Taking Users into
  Account: A Qualitative Evaluation of User Control Mechanisms in (News)
  Recommender Systems}. In \bibinfo{booktitle}{\emph{Proceedings of the 13th
  ACM Conference on Recommender Systems}} \emph{(\bibinfo{series}{RecSys
  '19})}. \bibinfo{pages}{69--77}.
\newblock


\bibitem[Hariri et~al\mbox{.}(2015)]%
        {hariri2015adapting}
\bibfield{author}{\bibinfo{person}{Negar Hariri}, \bibinfo{person}{Bamshad
  Mobasher}, {and} \bibinfo{person}{Robin Burke}.}
  \bibinfo{year}{2015}\natexlab{}.
\newblock \showarticletitle{Adapting to User Preference Changes in Interactive
  Recommendation}. In \bibinfo{booktitle}{\emph{Proceedings of the 24th
  International Conference on Artificial Intelligence}}
  \emph{(\bibinfo{series}{IJCAI '15})}. \bibinfo{pages}{4268--4274}.
\newblock


\bibitem[Harman(1992)]%
        {tipster}
\bibfield{author}{\bibinfo{person}{Donna Harman}.}
  \bibinfo{year}{1992}\natexlab{}.
\newblock \showarticletitle{The DARPA Tipster Project}. In
  \bibinfo{booktitle}{\emph{ACM SIGIR Forum}}, Vol.~\bibinfo{volume}{26}.
  \bibinfo{pages}{26--28}.
\newblock


\bibitem[Harman(1993)]%
        {harman:trec-1-data-description}
\bibfield{author}{\bibinfo{person}{Donna Harman}.}
  \bibinfo{year}{1993}\natexlab{}.
\newblock \showarticletitle{Document Detection Data Preparation}. In
  \bibinfo{booktitle}{\emph{TIPSTER TEXT PROGRAM: PHASE {I}: Proceedings of a
  Workshop held at Fredricksburg, Virginia, September 19-23, 1993}}.
  \bibinfo{pages}{17--31}.
\newblock


\bibitem[Harper et~al\mbox{.}(2015)]%
        {Harper:2015:RecSys}
\bibfield{author}{\bibinfo{person}{F.~Maxwell Harper}, \bibinfo{person}{Funing
  Xu}, \bibinfo{person}{Harmanpreet Kaur}, \bibinfo{person}{Kyle Condiff},
  \bibinfo{person}{Shuo Chang}, {and} \bibinfo{person}{Loren Terveen}.}
  \bibinfo{year}{2015}\natexlab{}.
\newblock \showarticletitle{Putting Users in Control of Their Recommendations}.
  In \bibinfo{booktitle}{\emph{Proceedings of the 9th ACM Conference on
  Recommender Systems}} \emph{(\bibinfo{series}{RecSys '15})}.
  \bibinfo{pages}{3--10}.
\newblock


\bibitem[Hensley(1963)]%
        {hensley:sdi-review}
\bibfield{author}{\bibinfo{person}{C.~B. Hensley}.}
  \bibinfo{year}{1963}\natexlab{}.
\newblock \showarticletitle{Selective Dissemination of Information (SDI): State
  of the Art in May, 1963}. In \bibinfo{booktitle}{\emph{Proceedings of the May
  21-23, 1963, Spring Joint Computer Conference}} \emph{(\bibinfo{series}{AFIPS
  '63 (Spring)})}. \bibinfo{pages}{257--262}.
\newblock


\bibitem[Hijikata et~al\mbox{.}(2012)]%
        {Hijikata:2012:SAC}
\bibfield{author}{\bibinfo{person}{Yoshinori Hijikata}, \bibinfo{person}{Yuki
  Kai}, {and} \bibinfo{person}{Shogo Nishida}.}
  \bibinfo{year}{2012}\natexlab{}.
\newblock \showarticletitle{The Relation between User Intervention and User
  Satisfaction for Information Recommendation}. In
  \bibinfo{booktitle}{\emph{Proceedings of the 27th Annual ACM Symposium on
  Applied Computing}} \emph{(\bibinfo{series}{SAC '12})}.
  \bibinfo{pages}{2002--2007}.
\newblock


\bibitem[Jannach et~al\mbox{.}(2017)]%
        {Jannach:2017:EWT}
\bibfield{author}{\bibinfo{person}{Dietmar Jannach}, \bibinfo{person}{Sidra
  Naveed}, {and} \bibinfo{person}{Michael Jugovac}.}
  \bibinfo{year}{2017}\natexlab{}.
\newblock \showarticletitle{User Control in Recommender Systems: Overview and
  Interaction Challenges}. In \bibinfo{booktitle}{\emph{E-Commerce and Web
  Technologies - 17th International Conference, Revised Selected Papers}}
  \emph{(\bibinfo{series}{EC-Web '16})}. \bibinfo{pages}{21--33}.
\newblock


\bibitem[Jin et~al\mbox{.}(2018)]%
        {Jin:RecSys:2018}
\bibfield{author}{\bibinfo{person}{Yucheng Jin}, \bibinfo{person}{Nava
  Tintarev}, {and} \bibinfo{person}{Katrien Verbert}.}
  \bibinfo{year}{2018}\natexlab{}.
\newblock \showarticletitle{Effects of Personal Characteristics on Music
  Recommender Systems with Different Levels of Controllability}. In
  \bibinfo{booktitle}{\emph{Proceedings of the 12th ACM Conference on
  Recommender Systems}} \emph{(\bibinfo{series}{RecSys '18})}.
  \bibinfo{pages}{13--21}.
\newblock


\bibitem[Kim et~al\mbox{.}(2009)]%
        {kim:2009:triptip}
\bibfield{author}{\bibinfo{person}{Jinyoung Kim}, \bibinfo{person}{Hyungjin
  Kim}, {and} \bibinfo{person}{Jung-hee Ryu}.} \bibinfo{year}{2009}\natexlab{}.
\newblock \showarticletitle{TripTip: A Trip Planning Service with Tag-based
  Recommendation}. In \bibinfo{booktitle}{\emph{CHI'09 Extended Abstracts on
  Human Factors in Computing Systems}} \emph{(\bibinfo{series}{CHI EA '09})}.
  \bibinfo{pages}{3467--3472}.
\newblock


\bibitem[Knijnenburg et~al\mbox{.}(2012)]%
        {Knijnenburg:2012:RecSys}
\bibfield{author}{\bibinfo{person}{Bart~P. Knijnenburg},
  \bibinfo{person}{Svetlin Bostandjiev}, \bibinfo{person}{John O'Donovan},
  {and} \bibinfo{person}{Alfred Kobsa}.} \bibinfo{year}{2012}\natexlab{}.
\newblock \showarticletitle{Inspectability and Control in Social Recommenders}.
  In \bibinfo{booktitle}{\emph{Proceedings of the Sixth ACM Conference on
  Recommender Systems}} \emph{(\bibinfo{series}{RecSys '12})}.
  \bibinfo{pages}{43--50}.
\newblock


\bibitem[Kuhn and Eickhoff(2016)]%
        {kuhn:negative-feedback}
\bibfield{author}{\bibinfo{person}{Lorenz Kuhn} {and} \bibinfo{person}{Carsten
  Eickhoff}.} \bibinfo{year}{2016}\natexlab{}.
\newblock \showarticletitle{Implicit Negative Feedback in Clinical Information
  Retrieval}. In \bibinfo{booktitle}{\emph{Proceedings of the SIGIR 2016
  Medical Information Retrieval Workshop (MedIR)}}.
\newblock


\bibitem[Lam et~al\mbox{.}(2008)]%
        {10.1145/1352793.1352837}
\bibfield{author}{\bibinfo{person}{Xuan~Nhat Lam}, \bibinfo{person}{Thuc Vu},
  \bibinfo{person}{Trong~Duc Le}, {and} \bibinfo{person}{Anh~Duc Duong}.}
  \bibinfo{year}{2008}\natexlab{}.
\newblock \showarticletitle{Addressing Cold-Start Problem in Recommendation
  Systems}. In \bibinfo{booktitle}{\emph{Proceedings of the 2nd International
  Conference on Ubiquitous Information Management and Communication}}
  \emph{(\bibinfo{series}{ICUIMC '08})}. \bibinfo{pages}{208--211}.
\newblock


\bibitem[Lee et~al\mbox{.}(2019)]%
        {lee2019melu}
\bibfield{author}{\bibinfo{person}{Hoyeop Lee}, \bibinfo{person}{Jinbae Im},
  \bibinfo{person}{Seongwon Jang}, \bibinfo{person}{Hyunsouk Cho}, {and}
  \bibinfo{person}{Sehee Chung}.} \bibinfo{year}{2019}\natexlab{}.
\newblock \showarticletitle{MeLU: Meta-Learned User Preference Estimator for
  Cold-Start Recommendation}. In \bibinfo{booktitle}{\emph{Proceedings of the
  25th ACM SIGKDD International Conference on Knowledge Discovery \& Data
  Mining}} \emph{(\bibinfo{series}{KDD '19})}. \bibinfo{pages}{1073--1082}.
\newblock


\bibitem[Leqi et~al\mbox{.}(2021)]%
        {leqi:satiation}
\bibfield{author}{\bibinfo{person}{Liu Leqi}, \bibinfo{person}{Fatma
  Kilinc-Karzan}, \bibinfo{person}{Zachary~C. Lipton}, {and}
  \bibinfo{person}{Alan~L. Montgomery}.} \bibinfo{year}{2021}\natexlab{}.
\newblock \showarticletitle{Rebounding Bandits for Modeling Satiation Effects}.
  In \bibinfo{booktitle}{\emph{Advances in Neural Information Processing
  Systems 34 pre-proceedings}} \emph{(\bibinfo{series}{NeurIPS '21})}.
\newblock


\bibitem[Lester et~al\mbox{.}(2021)]%
        {prompt-tuning-lester-etal-2021-power}
\bibfield{author}{\bibinfo{person}{Brian Lester}, \bibinfo{person}{Rami
  Al-Rfou}, {and} \bibinfo{person}{Noah Constant}.}
  \bibinfo{year}{2021}\natexlab{}.
\newblock \showarticletitle{The Power of Scale for Parameter-Efficient Prompt
  Tuning}. In \bibinfo{booktitle}{\emph{Proceedings of the 2021 Conference on
  Empirical Methods in Natural Language Processing}}
  \emph{(\bibinfo{series}{EMNLP '21})}. \bibinfo{pages}{3045--3059}.
\newblock


\bibitem[Luhn(1958)]%
        {luhn:SDI}
\bibfield{author}{\bibinfo{person}{Hans~Peter Luhn}.}
  \bibinfo{year}{1958}\natexlab{}.
\newblock \showarticletitle{A Business Intelligence System}.
\newblock \bibinfo{journal}{\emph{IBM J. Res. Dev.}} \bibinfo{volume}{2},
  \bibinfo{number}{4} (\bibinfo{date}{Oct.} \bibinfo{year}{1958}),
  \bibinfo{pages}{314--319}.
\newblock


\bibitem[Lukoff et~al\mbox{.}(2021)]%
        {Lukoff:2021:CHI}
\bibfield{author}{\bibinfo{person}{Kai Lukoff}, \bibinfo{person}{Ulrik Lyngs},
  \bibinfo{person}{Himanshu Zade}, \bibinfo{person}{J.~Vera Liao},
  \bibinfo{person}{James Choi}, \bibinfo{person}{Kaiyue Fan},
  \bibinfo{person}{Sean~A. Munson}, {and} \bibinfo{person}{Alexis Hiniker}.}
  \bibinfo{year}{2021}\natexlab{}.
\newblock \showarticletitle{How the Design of YouTube Influences User Sense of
  Agency}. In \bibinfo{booktitle}{\emph{Proceedings of the 2021 CHI Conference
  on Human Factors in Computing Systems}} \emph{(\bibinfo{series}{CHI '21})}.
  Article \bibinfo{articleno}{368}.
\newblock


\bibitem[Lyu et~al\mbox{.}(2021)]%
        {lyu2021workflow}
\bibfield{author}{\bibinfo{person}{Shengnan Lyu}, \bibinfo{person}{Arpit Rana},
  \bibinfo{person}{Scott Sanner}, {and} \bibinfo{person}{Mohamed~Reda
  Bouadjenek}.} \bibinfo{year}{2021}\natexlab{}.
\newblock \showarticletitle{A Workflow Analysis of Context-driven
  Conversational Recommendation}. In \bibinfo{booktitle}{\emph{Proceedings of
  the Web Conference 2021}} \emph{(\bibinfo{series}{WWW '21})}.
  \bibinfo{pages}{866--877}.
\newblock


\bibitem[Matthijs and Radlinski(2011)]%
        {Matthijs:2011:Personalizing}
\bibfield{author}{\bibinfo{person}{Nicolaas Matthijs} {and}
  \bibinfo{person}{Filip Radlinski}.} \bibinfo{year}{2011}\natexlab{}.
\newblock \showarticletitle{Personalizing Web Search Using Long Term Browsing
  History}. In \bibinfo{booktitle}{\emph{Proceedings of the Fourth ACM
  International Conference on Web Search and Data Mining}}
  \emph{(\bibinfo{series}{WSDM '11})}. \bibinfo{pages}{25--34}.
\newblock


\bibitem[McNee et~al\mbox{.}(2006)]%
        {mcnee:hci-recsys}
\bibfield{author}{\bibinfo{person}{Sean~M. McNee}, \bibinfo{person}{John
  Riedl}, {and} \bibinfo{person}{Joseph~A. Konstan}.}
  \bibinfo{year}{2006}\natexlab{}.
\newblock \showarticletitle{Making Recommendations Better: An Analytic Model
  for Human-Recommender Interaction}. In \bibinfo{booktitle}{\emph{CHI '06
  Extended Abstracts on Human Factors in Computing Systems}}
  \emph{(\bibinfo{series}{CHI EA '06})}. \bibinfo{pages}{1103--1108}.
\newblock


\bibitem[Narang et~al\mbox{.}(2020)]%
        {Narang2020WT5TT}
\bibfield{author}{\bibinfo{person}{Sharan Narang}, \bibinfo{person}{Colin
  Raffel}, \bibinfo{person}{Katherine Lee}, \bibinfo{person}{Adam Roberts},
  \bibinfo{person}{Noah Fiedel}, {and} \bibinfo{person}{Karishma Malkan}.}
  \bibinfo{year}{2020}\natexlab{}.
\newblock \bibinfo{title}{WT5?! Training Text-to-Text Models to Explain their
  Predictions}.
\newblock
\newblock
\showeprint[arxiv]{2004.14546}~[cs.CL]


\bibitem[Nema et~al\mbox{.}(2021)]%
        {Nema:2021:Disentangling}
\bibfield{author}{\bibinfo{person}{Preksha Nema}, \bibinfo{person}{Alexandros
  Karatzoglou}, {and} \bibinfo{person}{Filip Radlinski}.}
  \bibinfo{year}{2021}\natexlab{}.
\newblock \showarticletitle{Disentangling Preference Representations for
  Recommendation Critiquing with $\beta-$VAE}. In
  \bibinfo{booktitle}{\emph{Proceedings of the 30th ACM International
  Conference on Information \& Knowledge Management}}
  \emph{(\bibinfo{series}{CIKM '21})}. \bibinfo{pages}{1356--1365}.
\newblock


\bibitem[Ngo et~al\mbox{.}(2020)]%
        {Ngo:2020:UMAP}
\bibfield{author}{\bibinfo{person}{Thao Ngo}, \bibinfo{person}{Johannes
  Kunkel}, {and} \bibinfo{person}{J\"{u}rgen Ziegler}.}
  \bibinfo{year}{2020}\natexlab{}.
\newblock \showarticletitle{Exploring Mental Models for Transparent and
  Controllable Recommender Systems: A Qualitative Study}. In
  \bibinfo{booktitle}{\emph{Proceedings of the 28th ACM Conference on User
  Modeling, Adaptation and Personalization}} \emph{(\bibinfo{series}{UMAP
  '20})}. \bibinfo{pages}{183--191}.
\newblock


\bibitem[O'Donovan and Smyth(2005)]%
        {odonovan:2005:trust}
\bibfield{author}{\bibinfo{person}{John O'Donovan} {and} \bibinfo{person}{Barry
  Smyth}.} \bibinfo{year}{2005}\natexlab{}.
\newblock \showarticletitle{Trust in Recommender Systems}. In
  \bibinfo{booktitle}{\emph{Proceedings of the 10th International Conference on
  Intelligent User Interfaces}} \emph{(\bibinfo{series}{IUI '05})}.
  \bibinfo{pages}{167--174}.
\newblock


\bibitem[Paik and Oard(2014)]%
        {Paik:2014:CIKM}
\bibfield{author}{\bibinfo{person}{Jiaul~H. Paik} {and}
  \bibinfo{person}{Douglas~W. Oard}.} \bibinfo{year}{2014}\natexlab{}.
\newblock \showarticletitle{A Fixed-Point Method for Weighting Terms in Verbose
  Informational Queries}. In \bibinfo{booktitle}{\emph{Proceedings of the 23rd
  ACM International Conference on Conference on Information and Knowledge
  Management}} \emph{(\bibinfo{series}{CIKM '14})}. \bibinfo{pages}{131--140}.
\newblock


\bibitem[Parapar and Radlinski(2021a)]%
        {Parapar:2021:DiversePreferences}
\bibfield{author}{\bibinfo{person}{Javier Parapar} {and} \bibinfo{person}{Filip
  Radlinski}.} \bibinfo{year}{2021}\natexlab{a}.
\newblock \showarticletitle{Diverse User Preference Elicitation with
  Multi-Armed Bandits}. In \bibinfo{booktitle}{\emph{Proceedings of the 14th
  ACM International Conference on Web Search and Data Mining}}
  \emph{(\bibinfo{series}{WSDM '21})}. \bibinfo{pages}{130--138}.
\newblock


\bibitem[Parapar and Radlinski(2021b)]%
        {parapar:2021:diversity}
\bibfield{author}{\bibinfo{person}{Javier Parapar} {and} \bibinfo{person}{Filip
  Radlinski}.} \bibinfo{year}{2021}\natexlab{b}.
\newblock \showarticletitle{Towards Unified Metrics for Accuracy and Diversity
  for Recommender Systems}. In \bibinfo{booktitle}{\emph{Proceedings of the
  Fifteenth ACM Conference on Recommender Systems}}
  \emph{(\bibinfo{series}{RecSys '21})}. \bibinfo{pages}{75--84}.
\newblock


\bibitem[Parra and Brusilovsky(2015)]%
        {Parra:2015:IJHCS}
\bibfield{author}{\bibinfo{person}{Denis Parra} {and} \bibinfo{person}{Peter
  Brusilovsky}.} \bibinfo{year}{2015}\natexlab{}.
\newblock \showarticletitle{User-controllable Personalization: A Case Study
  with SetFusion}.
\newblock \bibinfo{journal}{\emph{Int. J. Hum. Comput.}}  \bibinfo{volume}{78}
  (\bibinfo{year}{2015}), \bibinfo{pages}{43--67}.
\newblock


\bibitem[Parra-Arnau et~al\mbox{.}(2017)]%
        {parra:2017:myadchoices}
\bibfield{author}{\bibinfo{person}{Javier Parra-Arnau},
  \bibinfo{person}{Jagdish~Prasad Achara}, {and} \bibinfo{person}{Claude
  Castelluccia}.} \bibinfo{year}{2017}\natexlab{}.
\newblock \showarticletitle{MyAdChoices: Bringing Transparency and Control to
  Online Advertising}.
\newblock \bibinfo{journal}{\emph{ACM Trans. Web}} \bibinfo{volume}{11},
  \bibinfo{number}{1} (\bibinfo{year}{2017}), \bibinfo{pages}{1--47}.
\newblock


\bibitem[Pereira et~al\mbox{.}(2018)]%
        {pereira:2018:preference-dynamics}
\bibfield{author}{\bibinfo{person}{Fabíola S.~F. Pereira},
  \bibinfo{person}{João Gama}, \bibinfo{person}{Sandra de Amo}, {and}
  \bibinfo{person}{Gina M.~B. Oliveira}.} \bibinfo{year}{2018}\natexlab{}.
\newblock \showarticletitle{On Analyzing User Preference Dynamics with Temporal
  Social Networks}.
\newblock \bibinfo{journal}{\emph{Mach. Learn.}} \bibinfo{number}{107}
  (\bibinfo{year}{2018}), \bibinfo{pages}{1745--1773}.
\newblock


\bibitem[Radlinski et~al\mbox{.}(2019)]%
        {Radlinski:2019:CCPE}
\bibfield{author}{\bibinfo{person}{Filip Radlinski}, \bibinfo{person}{Krisztian
  Balog}, \bibinfo{person}{Bill Byrne}, {and} \bibinfo{person}{Karthik
  Krishnamoorthi}.} \bibinfo{year}{2019}\natexlab{}.
\newblock \showarticletitle{Coached Conversational Preference Elicitation: A
  Case Study in Understanding Movie Preferences}. In
  \bibinfo{booktitle}{\emph{Proceedings of the Annual SIGdial Meeting on
  Discourse and Dialogue}} \emph{(\bibinfo{series}{SIGDIAL '19})}.
  \bibinfo{pages}{353--360}.
\newblock


\bibitem[Radlinski et~al\mbox{.}(2022)]%
        {Radlinski:2022:Subjectivity}
\bibfield{author}{\bibinfo{person}{Filip Radlinski}, \bibinfo{person}{Craig
  Boutilier}, \bibinfo{person}{Deepak Ramachandran}, {and}
  \bibinfo{person}{Ivan Vendrov}.} \bibinfo{year}{2022}\natexlab{}.
\newblock \showarticletitle{Subjective Attributes in Conversational
  Recommendation Systems: Challenges and Opportunities}. In
  \bibinfo{booktitle}{\emph{Proceedings of the 36th AAAI Conference on
  Artificial Intelligence}} \emph{(\bibinfo{series}{AAAI '22})}.
\newblock


\bibitem[Rago et~al\mbox{.}(2021)]%
        {rago:2021:argumentative}
\bibfield{author}{\bibinfo{person}{Antonio Rago}, \bibinfo{person}{Oana
  Cocarascu}, \bibinfo{person}{Christos Bechlivanidis}, \bibinfo{person}{David
  Lagnado}, {and} \bibinfo{person}{Francesca Toni}.}
  \bibinfo{year}{2021}\natexlab{}.
\newblock \showarticletitle{Argumentative Explanations for Interactive
  Recommendations}.
\newblock \bibinfo{journal}{\emph{Artif. Intell.}}  \bibinfo{volume}{296}
  (\bibinfo{year}{2021}), \bibinfo{pages}{103506}.
\newblock


\bibitem[Ribeiro et~al\mbox{.}(2020)]%
        {ribeiro:2020:investigating}
\bibfield{author}{\bibinfo{person}{Leonardo~FR Ribeiro},
  \bibinfo{person}{Martin Schmitt}, \bibinfo{person}{Hinrich Sch{\"u}tze},
  {and} \bibinfo{person}{Iryna Gurevych}.} \bibinfo{year}{2020}\natexlab{}.
\newblock \bibinfo{title}{Investigating Pretrained Language Models for
  Graph-to-Text Generation}.
\newblock
\newblock
\showeprint[arxiv]{2007.08426}~[cs.CL]


\bibitem[Ricci et~al\mbox{.}(2015)]%
        {recsys-handbook}
\bibfield{author}{\bibinfo{person}{Francesco Ricci}, \bibinfo{person}{Lior
  Rokach}, {and} \bibinfo{person}{Bracha Shapira}.}
  \bibinfo{year}{2015}\natexlab{}.
\newblock \bibinfo{booktitle}{\emph{Recommender Systems Handbook}
  (\bibinfo{edition}{2nd} ed.)}.
\newblock \bibinfo{publisher}{Springer}.
\newblock


\bibitem[Sachdeva and McAuley(2020)]%
        {sachdeva2020usefulReview}
\bibfield{author}{\bibinfo{person}{Noveen Sachdeva} {and}
  \bibinfo{person}{Julian McAuley}.} \bibinfo{year}{2020}\natexlab{}.
\newblock \showarticletitle{How Useful Are Reviews for Recommendation? {A}
  Critical Review and Potential Improvements}. In
  \bibinfo{booktitle}{\emph{Proceedings of the 43rd International ACM SIGIR
  Conference on Research and Development in Information Retrieval}}
  \emph{(\bibinfo{series}{SIGIR '20})}. \bibinfo{pages}{1845--1848}.
\newblock


\bibitem[Sen et~al\mbox{.}(2007)]%
        {Sen:2007:GROUP}
\bibfield{author}{\bibinfo{person}{Shilad Sen}, \bibinfo{person}{F.~Maxwell
  Harper}, \bibinfo{person}{Adam LaPitz}, {and} \bibinfo{person}{John Riedl}.}
  \bibinfo{year}{2007}\natexlab{}.
\newblock \showarticletitle{The Quest for Quality Tags}. In
  \bibinfo{booktitle}{\emph{Proceedings of the 2007 International ACM
  Conference on Supporting Group Work}} \emph{(\bibinfo{series}{GROUP '07})}.
  \bibinfo{pages}{361--370}.
\newblock


\bibitem[Sen et~al\mbox{.}(2006)]%
        {Sen:2006:CSCW}
\bibfield{author}{\bibinfo{person}{Shilad Sen}, \bibinfo{person}{Shyong~K.
  Lam}, \bibinfo{person}{Al~Mamunur Rashid}, \bibinfo{person}{Dan Cosley},
  \bibinfo{person}{Dan Frankowski}, \bibinfo{person}{Jeremy Osterhouse},
  \bibinfo{person}{F.~Maxwell Harper}, {and} \bibinfo{person}{John Riedl}.}
  \bibinfo{year}{2006}\natexlab{}.
\newblock \showarticletitle{Tagging, Communities, Vocabulary, Evolution}. In
  \bibinfo{booktitle}{\emph{Proceedings of the 2006 20th Anniversary Conference
  on Computer Supported Cooperative Work}} \emph{(\bibinfo{series}{CSCW '06})}.
  \bibinfo{pages}{181--190}.
\newblock


\bibitem[Sinha and Swearingen(2002)]%
        {Sinha:2002:CHI}
\bibfield{author}{\bibinfo{person}{Rashmi Sinha} {and} \bibinfo{person}{Kirsten
  Swearingen}.} \bibinfo{year}{2002}\natexlab{}.
\newblock \showarticletitle{The Role of Transparency in Recommender Systems}.
  In \bibinfo{booktitle}{\emph{CHI '02 Extended Abstracts on Human Factors in
  Computing Systems}} \emph{(\bibinfo{series}{CHI EA '02})}.
  \bibinfo{pages}{830--831}.
\newblock


\bibitem[Stefanidis et~al\mbox{.}(2011)]%
        {Stefanidis:2011:contextual-preferences}
\bibfield{author}{\bibinfo{person}{Kostas Stefanidis},
  \bibinfo{person}{Evaggelia Pitoura}, {and} \bibinfo{person}{Panos
  Vassiliadis}.} \bibinfo{year}{2011}\natexlab{}.
\newblock \showarticletitle{Managing Contextual Preferences}.
\newblock \bibinfo{journal}{\emph{Inf. Syst.}} \bibinfo{volume}{36},
  \bibinfo{number}{8} (\bibinfo{year}{2011}), \bibinfo{pages}{1158--1180}.
\newblock
\showISSN{0306-4379}


\bibitem[Sun et~al\mbox{.}(2020)]%
        {sun:2020:where}
\bibfield{author}{\bibinfo{person}{Ke Sun}, \bibinfo{person}{Tieyun Qian},
  \bibinfo{person}{Tong Chen}, \bibinfo{person}{Yile Liang},
  \bibinfo{person}{Quoc Viet~Hung Nguyen}, {and} \bibinfo{person}{Hongzhi
  Yin}.} \bibinfo{year}{2020}\natexlab{}.
\newblock \showarticletitle{Where to Go Next: Modeling Long- and Short-Term
  User Preferences for Point-of-Interest Recommendation}.
\newblock \bibinfo{journal}{\emph{Proceedings of the AAAI Conference on
  Artificial Intelligence}} \bibinfo{volume}{34}, \bibinfo{number}{01}
  (\bibinfo{year}{2020}), \bibinfo{pages}{214--221}.
\newblock


\bibitem[Sun and Zhang(2018)]%
        {Sun:2018:Conversational}
\bibfield{author}{\bibinfo{person}{Yueming Sun} {and} \bibinfo{person}{Yi
  Zhang}.} \bibinfo{year}{2018}\natexlab{}.
\newblock \showarticletitle{Conversational Recommender System}. In
  \bibinfo{booktitle}{\emph{Proceedings of the 41st International ACM SIGIR
  Conference on Research and Development in Information Retrieval}}
  \emph{(\bibinfo{series}{SIGIR '18})}. \bibinfo{pages}{235--244}.
\newblock


\bibitem[Teevan et~al\mbox{.}(2005)]%
        {Teevan:2005:Personalizing}
\bibfield{author}{\bibinfo{person}{Jaime Teevan}, \bibinfo{person}{Susan~T.
  Dumais}, {and} \bibinfo{person}{Eric Horvitz}.}
  \bibinfo{year}{2005}\natexlab{}.
\newblock \showarticletitle{Personalizing Search via Automated Analysis of
  Interests and Activities}. In \bibinfo{booktitle}{\emph{Proceedings of the
  International ACM SIGIR Conference on Research and Development in Information
  Retrieval}} \emph{(\bibinfo{series}{SIGIR '05})}. \bibinfo{pages}{235--244}.
\newblock


\bibitem[Thoppilan et~al\mbox{.}(2022)]%
        {thoppilan:2022:lamda}
\bibfield{author}{\bibinfo{person}{Romal Thoppilan}, \bibinfo{person}{Daniel~De
  Freitas}, \bibinfo{person}{Jamie Hall}, \bibinfo{person}{Noam Shazeer},
  \bibinfo{person}{Apoorv Kulshreshtha}, \bibinfo{person}{Heng-Tze Cheng},
  \bibinfo{person}{Alicia Jin}, \bibinfo{person}{Taylor Bos},
  \bibinfo{person}{Leslie Baker}, \bibinfo{person}{Yu Du},
  \bibinfo{person}{YaGuang Li}, \bibinfo{person}{Hongrae Lee},
  \bibinfo{person}{Huaixiu~Steven Zheng}, \bibinfo{person}{Amin Ghafouri},
  \bibinfo{person}{Marcelo Menegali}, \bibinfo{person}{Yanping Huang},
  \bibinfo{person}{Maxim Krikun}, \bibinfo{person}{Dmitry Lepikhin},
  \bibinfo{person}{James Qin}, \bibinfo{person}{Dehao Chen},
  \bibinfo{person}{Yuanzhong Xu}, \bibinfo{person}{Zhifeng Chen},
  \bibinfo{person}{Adam Roberts}, \bibinfo{person}{Maarten Bosma},
  \bibinfo{person}{Yanqi Zhou}, \bibinfo{person}{Chung-Ching Chang},
  \bibinfo{person}{Igor Krivokon}, \bibinfo{person}{Will Rusch},
  \bibinfo{person}{Marc Pickett}, \bibinfo{person}{Kathleen Meier-Hellstern},
  \bibinfo{person}{Meredith~Ringel Morris}, \bibinfo{person}{Tulsee Doshi},
  \bibinfo{person}{Renelito~Delos Santos}, \bibinfo{person}{Toju Duke},
  \bibinfo{person}{Johnny Soraker}, \bibinfo{person}{Ben Zevenbergen},
  \bibinfo{person}{Vinodkumar Prabhakaran}, \bibinfo{person}{Mark Diaz},
  \bibinfo{person}{Ben Hutchinson}, \bibinfo{person}{Kristen Olson},
  \bibinfo{person}{Alejandra Molina}, \bibinfo{person}{Erin Hoffman-John},
  \bibinfo{person}{Josh Lee}, \bibinfo{person}{Lora Aroyo},
  \bibinfo{person}{Ravi Rajakumar}, \bibinfo{person}{Alena Butryna},
  \bibinfo{person}{Matthew Lamm}, \bibinfo{person}{Viktoriya Kuzmina},
  \bibinfo{person}{Joe Fenton}, \bibinfo{person}{Aaron Cohen},
  \bibinfo{person}{Rachel Bernstein}, \bibinfo{person}{Ray Kurzweil},
  \bibinfo{person}{Blaise Aguera-Arcas}, \bibinfo{person}{Claire Cui},
  \bibinfo{person}{Marian Croak}, \bibinfo{person}{Ed Chi}, {and}
  \bibinfo{person}{Quoc Le}.} \bibinfo{year}{2022}\natexlab{}.
\newblock \bibinfo{title}{LaMDA: Language Models for Dialog Applications}.
\newblock
\newblock
\showeprint[arxiv]{2201.08239}~[cs.CL]


\bibitem[Tintarev and Masthoff(2012)]%
        {Tintarev:2012:UMUAP}
\bibfield{author}{\bibinfo{person}{Nava Tintarev} {and} \bibinfo{person}{Judith
  Masthoff}.} \bibinfo{year}{2012}\natexlab{}.
\newblock \showarticletitle{Evaluating the Effectiveness of Explanations for
  Recommender Systems}.
\newblock \bibinfo{journal}{\emph{User Model. User-adapt. Interact.}}
  \bibinfo{volume}{22} (\bibinfo{date}{oct} \bibinfo{year}{2012}),
  \bibinfo{pages}{399--439}.
\newblock


\bibitem[Tintarev and Masthoff(2015)]%
        {Tintarev:2015:ERD}
\bibfield{author}{\bibinfo{person}{Nava Tintarev} {and} \bibinfo{person}{Judith
  Masthoff}.} \bibinfo{year}{2015}\natexlab{}.
\newblock \showarticletitle{Explaining Recommendations: {D}esign and
  Evaluation}.
\newblock In \bibinfo{booktitle}{\emph{Recommender Systems Handbook}
  (\bibinfo{edition}{2nd} ed.)}, \bibfield{editor}{\bibinfo{person}{Francesco
  Ricci}, \bibinfo{person}{Lior Rokach}, \bibinfo{person}{Bracha Shapira},
  {and} \bibinfo{person}{Paul~B. Kantor}} (Eds.).
  \bibinfo{publisher}{Springer}, \bibinfo{pages}{353--382}.
\newblock


\bibitem[Trienes and Balog(2019)]%
        {Trienes:2019:IUQ}
\bibfield{author}{\bibinfo{person}{Jan Trienes} {and}
  \bibinfo{person}{Krisztian Balog}.} \bibinfo{year}{2019}\natexlab{}.
\newblock \showarticletitle{Identifying Unclear Questions in Community Question
  Answering Websites}. In \bibinfo{booktitle}{\emph{Proceedings of the 41th
  European Conference on Advances in Information Retrieval}}
  \emph{(\bibinfo{series}{ECIR '19})}. \bibinfo{pages}{276--289}.
\newblock


\bibitem[Valcarce et~al\mbox{.}(2018)]%
        {Valcarce:2018:RecSys}
\bibfield{author}{\bibinfo{person}{Daniel Valcarce}, \bibinfo{person}{Alejandro
  Bellog\'{\i}n}, \bibinfo{person}{Javier Parapar}, {and}
  \bibinfo{person}{Pablo Castells}.} \bibinfo{year}{2018}\natexlab{}.
\newblock \showarticletitle{On the Robustness and Discriminative Power of
  Information Retrieval Metrics for Top-N Recommendation}. In
  \bibinfo{booktitle}{\emph{Proceedings of the 12th ACM Conference on
  Recommender Systems}} \emph{(\bibinfo{series}{RecSys '18})}.
  \bibinfo{pages}{260--268}.
\newblock


\bibitem[Vig et~al\mbox{.}(2009)]%
        {Vig:2009:tagsplanations}
\bibfield{author}{\bibinfo{person}{Jesse Vig}, \bibinfo{person}{Shilad Sen},
  {and} \bibinfo{person}{John Riedl}.} \bibinfo{year}{2009}\natexlab{}.
\newblock \showarticletitle{Tagsplanations: Explaining Recommendations Using
  Tags}. In \bibinfo{booktitle}{\emph{Proceedings of the 14th International
  Conference on Intelligent User Interfaces}} \emph{(\bibinfo{series}{IUI
  '09})}. \bibinfo{pages}{47--56}.
\newblock


\bibitem[Voorhees(2004)]%
        {ROBUST2004}
\bibfield{author}{\bibinfo{person}{Ellen~M. Voorhees}.}
  \bibinfo{year}{2004}\natexlab{}.
\newblock \showarticletitle{Overview of the {TREC} 2004 Robust Track}. In
  \bibinfo{booktitle}{\emph{Proceedings of the 13th Text REtrieval Conference
  (TREC 2004)}}.
\newblock


\bibitem[Voorhees and Harman(2005)]%
        {voorhees:trecbook}
\bibfield{editor}{\bibinfo{person}{Ellen~M. Voorhees} {and}
  \bibinfo{person}{Donna~K. Harman}} (Eds.). \bibinfo{year}{2005}\natexlab{}.
\newblock \bibinfo{booktitle}{\emph{TREC: Experiment and Evaluation in
  Information Retrieval}}.
\newblock \bibinfo{publisher}{{MIT} Press}.
\newblock


\bibitem[Wagner et~al\mbox{.}(2014)]%
        {Wagner:2014:WWW}
\bibfield{author}{\bibinfo{person}{Claudia Wagner}, \bibinfo{person}{Philipp
  Singer}, \bibinfo{person}{Markus Strohmaier}, {and}
  \bibinfo{person}{Bernardo~A. Huberman}.} \bibinfo{year}{2014}\natexlab{}.
\newblock \showarticletitle{Semantic Stability in Social Tagging Streams}. In
  \bibinfo{booktitle}{\emph{Proceedings of the 23rd International Conference on
  World Wide Web}} \emph{(\bibinfo{series}{WWW '14})}.
  \bibinfo{pages}{735--746}.
\newblock


\bibitem[Yang et~al\mbox{.}(2014)]%
        {Yang:2014:ECIR}
\bibfield{author}{\bibinfo{person}{Bishan Yang}, \bibinfo{person}{Nish Parikh},
  \bibinfo{person}{Gyanit Singh}, {and} \bibinfo{person}{Neel Sundaresan}.}
  \bibinfo{year}{2014}\natexlab{}.
\newblock \showarticletitle{A Study of Query Term Deletion Using Large-Scale
  E-commerce Search Logs}. In \bibinfo{booktitle}{\emph{Proceedings of the 36th
  European Conference on Information Retrieval Research: Advances in
  Information Retrieval}} \emph{(\bibinfo{series}{ECIR '14})}.
  \bibinfo{pages}{235--246}.
\newblock


\bibitem[Zamani et~al\mbox{.}(2022)]%
        {Zamani:2022:FnTIR}
\bibfield{author}{\bibinfo{person}{Hamed Zamani}, \bibinfo{person}{Johanne~R.
  Trippas}, \bibinfo{person}{Jeff Dalton}, {and} \bibinfo{person}{Filip
  Radlinski}.} \bibinfo{year}{2022}\natexlab{}.
\newblock \bibinfo{title}{Conversational Information Seeking}.
\newblock
\newblock
\showeprint[arxiv]{2201.08808}~[cs.IR]


\bibitem[Zemlyanskiy et~al\mbox{.}(2021)]%
        {doc-ent-2020}
\bibfield{author}{\bibinfo{person}{Yury Zemlyanskiy}, \bibinfo{person}{Sudeep
  Gandhe}, \bibinfo{person}{Ruining He}, \bibinfo{person}{Bhargav Kanagal},
  \bibinfo{person}{Anirudh Ravula}, \bibinfo{person}{Juro Gottweis},
  \bibinfo{person}{Fei Sha}, {and} \bibinfo{person}{Ilya Eckstein}.}
  \bibinfo{year}{2021}\natexlab{}.
\newblock \showarticletitle{DOCENT: Learning Self-Supervised Entity
  Representations from Large Document Collections}. In
  \bibinfo{booktitle}{\emph{Proceedings of the 16th Conference of the European
  Chapter of the Association for Computational Linguistics: Main Volume}}
  \emph{(\bibinfo{series}{EACL '21})}. \bibinfo{pages}{2540--2549}.
\newblock


\bibitem[Zhai and Lafferty(2004)]%
        {Zhai:2004:TOIS}
\bibfield{author}{\bibinfo{person}{Chengxiang Zhai} {and} \bibinfo{person}{John
  Lafferty}.} \bibinfo{year}{2004}\natexlab{}.
\newblock \showarticletitle{A Study of Smoothing Methods for Language Models
  Applied to Information Retrieval}.
\newblock \bibinfo{journal}{\emph{ACM Trans. Inf. Syst.}} \bibinfo{volume}{22},
  \bibinfo{number}{2} (\bibinfo{date}{apr} \bibinfo{year}{2004}),
  \bibinfo{pages}{179--214}.
\newblock


\bibitem[Zhang and Chen(2020)]%
        {zhang:explainable-recsys}
\bibfield{author}{\bibinfo{person}{Yongfeng Zhang} {and} \bibinfo{person}{Xu
  Chen}.} \bibinfo{year}{2020}\natexlab{}.
\newblock \showarticletitle{Explainable Recommendation: A Survey and New
  Perspectives}.
\newblock \bibinfo{journal}{\emph{Found. Trends Inf. Ret.}}
  \bibinfo{volume}{14}, \bibinfo{number}{1} (\bibinfo{year}{2020}),
  \bibinfo{pages}{1--101}.
\newblock


\bibitem[Zhang et~al\mbox{.}(2012)]%
        {zhang:2012:auralist}
\bibfield{author}{\bibinfo{person}{Yuan~Cao Zhang},
  \bibinfo{person}{Diarmuid~\'{O} S\'{e}aghdha}, \bibinfo{person}{Daniele
  Quercia}, {and} \bibinfo{person}{Tamas Jambor}.}
  \bibinfo{year}{2012}\natexlab{}.
\newblock \showarticletitle{Auralist: Introducing Serendipity into Music
  Recommendation}. In \bibinfo{booktitle}{\emph{Proceedings of the Fifth ACM
  International Conference on Web Search and Data Mining}}
  \emph{(\bibinfo{series}{WSDM '12})}. \bibinfo{pages}{13--22}.
\newblock


\bibitem[Zhao et~al\mbox{.}(2018)]%
        {zhao-etal-2018-gender}
\bibfield{author}{\bibinfo{person}{Jieyu Zhao}, \bibinfo{person}{Tianlu Wang},
  \bibinfo{person}{Mark Yatskar}, \bibinfo{person}{Vicente Ordonez}, {and}
  \bibinfo{person}{Kai-Wei Chang}.} \bibinfo{year}{2018}\natexlab{}.
\newblock \showarticletitle{Gender Bias in Coreference Resolution: Evaluation
  and Debiasing Methods}. In \bibinfo{booktitle}{\emph{Proceedings of the
  Conference of the North {A}merican Chapter of the Association for
  Computational Linguistics: Human Language Technologies, Volume 2 (Short
  Papers)}} \emph{(\bibinfo{series}{NAACL '18})}. \bibinfo{pages}{15--20}.
\newblock


\bibitem[Zhao and Callan(2010)]%
        {Zhao:2010:CIKM}
\bibfield{author}{\bibinfo{person}{Le Zhao} {and} \bibinfo{person}{Jamie
  Callan}.} \bibinfo{year}{2010}\natexlab{}.
\newblock \showarticletitle{Term Necessity Prediction}. In
  \bibinfo{booktitle}{\emph{Proceedings of the ACM International Conference on
  Information and Knowledge Management}} \emph{(\bibinfo{series}{CIKM '10})}.
  \bibinfo{pages}{259--268}.
\newblock


\end{thebibliography}

\end{document}